\documentstyle{mn}

   \title[GEMINI GMOS-IFU spectroscopy of BALs I. Mrk 231]
         {Deep GEMINI GMOS-IFU spectroscopy of BAL QSOs: \\
         I. Decoupling the BAL, QSO, starburst, NLR,  \\
         supergiant bubbles and galactic wind in Mrk 231
}

\author[Lipari et al.]
   {S. Lipari$^{1}$, S.F. Sanchez$^{2}$, M. Bergmann$^{3}$, R. Terlevich$^{4,5}$,
   B. Garcia$^{6}$, B. Punsly$^{7}$,   
   \newauthor
E. Mediavilla$^{6}$, Y. Taniguchi$^{8}$, M. Ajiki$^{8}$, W. Zheng$^{9}$,
J. Acosta$^{6}$, K. Jahnke$^{10}$ \\
$^{1}$ C\'ordoba Observatory and CONICET, Laprida 854, 5000 C\'ordoba, Argentina.\\
$^{2}$ Calar Alto Observatory, C/Jesus Durban Remon 2-2, E-04004 Almeria, Spain.\\
$^{3}$ Gemini Observatory, La Serena, Chile.\\
$^{4}$ Institute of Astronomy, Madingley Road, Cambridge CB3 OHA, UK. \\
$^{5}$ Instituto Nacional de Astrofisica Optica y Electronica (INAOE), Puebla, Mexico.\\
$^{6}$ Instituto de Astrofisica de Canarias, 38205 La Laguna, Tenerife, Spain.\\
$^{7}$ Centre for Relativistic Astrophysics, Univ. of Rome La Sapienza, Italy and USA.\\
$^{8}$ Astronomical Institute, Tohoku University, Aoba, Sendai 980--8578, Japan.\\
$^{9}$ Depart. of Physics and Astronomy, John Hopkins Univ., Baltimore, MD 21218, USA.\\
$^{10}$ Astrophysics Institute of Potsdam, An der Sternwarte 16, 14482 Potsdam, Germany.
}

\date{Received     ;
      in original form }

\pagerange{\pageref{firstpage}--\pageref{lastpage}}
\pubyear{2005}

\begin{document}

\maketitle

\label{firstpage}

\begin{abstract}

In this work we present the first results of a study of BAL QSOs
(at low and high redshift), based on very deep Gemini GMOS integral
field  spectroscopy.
In particular, the results obtained for the nearest BAL IR--QSO Mrk 231 are
presented.

For the nuclear region of Mrk 231, the QSO and host-galaxy components
were modelled, using a new technique of decoupling 3D spectra.
From this study, the following main results were found:
(i) in the pure host galaxy spectrum an extreme nuclear starburst
component was clearly observed,
mainly as a very strong increase in the flux, at the blue wavelengths;
(ii) the BAL system {\sc i}  is observed in the spectrum of the host galaxy;
(iii) in the clean/pure QSO emission spectrum, only broad lines were detected.
3D GMOS individual spectra (specially in the near IR Ca {\sc ii} triplet) and maps
confirm the presence of an extreme and young nuclear starburst (8 $<$ age $<$
15 Myr), which was detected mainly in a ring or toroid  with a radius
r $=$ 0.3$'' \sim$ 200 pc, around the very nucleus. The extreme
continuum blue component  was detected only to the
south of the very nucleus. This area is coincident with the region where
we previously suggested that the galactic wind is cleaning the nuclear dust.

Very deep three-dimensional (3D) spectra and maps clearly show that
the BAL systems {\sc i} and {\sc ii} --mainly in the strong {\it ``absorption lines"}
Na ID$\lambda$5889-95 and Ca {\sc ii} K$\lambda$3933-- are extended
(reaching $\sim$1.4--1.6$''$ $\sim$1.2--1.3 kpc, from the nucleus)
and clearly elongated at the position angle (PA) close to the radio jet PA.
Which suggest that the BAL systems {\sc i} and {\sc ii} are ``both" associated
with the radio jet.

The physical properties of the four expanding nuclear bubbles were analysed,
using the GMOS 3D spectra and maps. In particular, we found
strong multiple LINER/OF emission line systems and  Wolf Rayet features 
in the main knots of the more external super bubble S1 (r $=$ 3.0 kpc).
The kinematics of these knots --and the internal bubbles--
suggest that they are associated with an area of rupture of
the shell S1 (at the south-west). In addition,
in the more internal super bubble S4 and close to the very nucleus
(for r $<$ 0.7$" \sim$ 0.6 kpc), two similar narrow emission
line systems were detected, with strong [S {\sc ii}] and [O {\sc i}] emission
and $\Delta$V $\sim$ --200 km s$^{-1}$. These results suggest that
 an important part of the nuclear NLR is generated by the
OF process and the associated low velocity ionizing shocks.

Finally, the nature of the composite BAL systems and  very extended OF
process  --of 50 kpc--  in Mrk 231 (and similar QSOs) are discussed.
In addition, the {\it ``composite hyper--wind scenario"} (already proposed
for BALs) is suggested for the origin of giant Ly$\alpha$ blobs. The
importance of study the end phases of Mrk 231, and similar evolving
elliptical galaxies and QSOs (i.e., galaxy remnants) is discussed.

\end{abstract}


\section{INTRODUCTION}\label{pro-intro1}

Theoretical models based on the hierarchical clustering scenario suggest
that the first generation (i.e., Population III) massive stars could be
born around z = 30 (0.5 Gyr after the Big-Bang) and the galactic systems
with masses higher than 10$^{10}$  M$_{\odot}$ could be assembled at z = 5-10.
There is increase evidence that galactic outflow (OF) and BAL systems play
a main role in the high redshift universe, at z $>$ 5 (Frye, Broadhurst,
Benitez 2002; Maiolino et al. 2003, 2004a,b; Lipari et al. 2005a,b,c). 
Thus, a main step for the study of QSO and galaxy  formation at high redshift
is to understand the extreme outflows and  BAL  processes in nearby
QSOs/galaxies because we can obtain unambiguous data since they are close and
bright enough to be observed in detail. Motivated by this, our group began a
first  program on investigations of BAL, out-flow (OF) and galactic winds (GW) in nearby
IR QSOs and mergers (Lipari et al. 2005a,b,c, 2004a,b,c,d, 2003, 2000, 1994).

An evolutionary and composite scenario was proposed for BAL + IR +
Fe\,{\sc ii} QSOs (L\'{i}pari et al. 1993, 1994, 2003, 2005a; Lipari
\& Terlevich 2006; L\'{i}pari 1994).
Where mergers fuel extreme star formation
processes and  AGNs, resulting in strong dust and IR emission, large
number of SN and Hyper Nova events (probably in the accretion disks and/or
in the nuclear starburst ring/toroid) with expanding super giant
bubbles and shell. The BALs in IR + Fe\,{\sc ii} QSOs were associated with
this composite nature of the OF process.

Mrk 231 is the nearest BAL + IR + Fe II + GW QSO. Specifically,
it shows very interesting spectral characteristics, dominated
in the optical by extremely strong Fe {\sc ii} and broad Balmer emission
lines at Z$_{em} \sim$  0.042, plus remarkable absorption line systems.
In particular, Mrk 231 shows  two type of absorption line systems:
a clear stellar absorption at Z$_{abs}  \sim$ 0.042 plus at least three
strong broad absorption line (BAL) systems.
These strong BAL systems show the following velocity of ejection:
V$_{eject}$ of BAL I, {\sc ii} and {\sc iii} of $\sim$4700, $\sim$6000,
and $\sim$8000 km s$^{-1}$, respectively
(see for details and references Lipari et al. 2005a).

On the other hand,
Mrk 231 is one of the most luminous IR object in the local
universe, with L$_{IR[8-1000 \mu m]} = 3.56 \times 10^{12} L_{\odot}$,
L$_{IR}$/L$_{B}$ = 32, M$_K =$ --24.7
 and M$_V =$ --22.5 (Markarian 1969; Adams 1972; Rieke \& Low 1972, 1975;
Boksenberg et al. 1997; Cutri, Rieke \& Lebofsky 1984, and others).
The origin of this extreme IR luminosity  is associated with the
two main sources of nuclear energy: an AGN plus an extreme nuclear and
circumnuclear starburst (see Lipari et al. 1994, 2005a).

Throughout the paper, a Hubble constant of H$_{0}$ = 75 km~s$^{-1}$
Mpc$^{-1}$ will be assumed.
For Mrk\,231 we adopted the distance of $\sim$168 Mpc (from Lipari et al.
2005a and from this paper: Sections 3, 11 and 12).
This distance was obtained from the stellar
absorption lines and the main emission line component, with a final
value of redshift z = 0.04218 and cz = 12654 $\pm$10 km~s$^{-1}$.
Thus, the angular scale is 1$'' \approx$814 pc. \\


\section{GEMINI PROGRAMME OF BAL QSOs}\label{gpro2}

In this paper we present the first results from a new part of our
observational programme of BAL QSOs: a study of high spatial and spectral
resolution of 3D Gemini spectroscopy of nearby BAL + IR + Fe {\sc ii} QSOs,
selected from our database of IR QSOs/mergers outflow (Lipari et al.
2005a, 2004c).
We have observed 7 nearby and high z BAL QSOs, mainly from our original
sample of more than 50 IR QSOs/Mergers with OF, plus Submm + Radio SDSS-QSOs.

The general goal of this programme is to study the kinematics,
physical conditions and morphology of the gas and the stars in the very
nucleus of BAL + IR + Fe {\sc ii} QSOs.
Some of the particular goals of this programme can be enumerated as
follows:

\begin{enumerate}

\item
To study BALs associated with composite QSOs: AGNs + starburst (with jet +
accretions disks, shells).
Specifically, we are interested to study in detail and to expand
our finding that some BALs systems are linked with: (1) bipolar OF
probably generated by sub relativistic jets (Lipari et al. 2005a;
Punsly \& Lipari 2005); (2) supergiant explosive events, probably
associated with Hyper Novae (Lipari et al. 2005a).

Our purpose is to study the spatial distribution, and the
kinematics and physical properties of the ionized gas, stars and dust, all
of which provide information about the evolutionary conditions of the ISM
and the possible origins of extreme star formation and explosive processes.

\item
To investigate the presence, properties, origin and importance of galactic
winds, stellar populations and giant explosions in BAL + IR + Fe II
QSOs/mergers.

In particular,
it is important to study in detail the host galaxies of BAL + IR QSOs,
specially the presence of galactic winds, young stellar populations
(with Wolf--Rayet features, etc), SN, HyN, and giant arcs/shells.

\item
To study possible links or evolutionary paths among mergers,
starbursts, BAL, QSOs and galaxies. More specifically, to analyse the
possible connection between IR mergers with extreme starburst + GW leading
to IR composite QSOs with GW, and elliptical galaxies.

We have a special interest in studying the evolutionary role of IR QSOs with
BAL + giant galactic shells (plus strong Fe\ {\sc ii} emission).
These IR QSOs  were defined as {\it composite and transition} objects
between ULIRGs and standard QSOs, in the IRAS colour--colour diagram
(L\'{\i}pari 1994; L\'{\i}pari et al. 2005a: fig. 5 and 15, respectively);
e.g., Mrk 231, IRAS\,07598+6508, IRAS\,17002+5153, IRAS\,04505-2958 and others.

\item
To study the origin of low ionization BAL systems observed in IR QSOs
(Boroson \& Mayer 1992; L\'{\i}pari et al. 1994; L\'{\i}pari 1994).
These absorption systems can be explained within the framework of the
composite (starburst + AGN) outflow scenario. We are particularly interested
in making a detailed study of our proposition that some BAL systems and
extreme Fe {\sc ii} emission could be associated with galactic winds that
produce dusty expanding shells.

\end{enumerate}

A second program of this research is the study of OF/BAL in forming galaxies
and QSOs at high redshift (z $>$ 2).
We are studying very deep 3D spectroscopic data of 
Sub-mm and Radio SDSS BAL-QSOs, using Gemini+GMOS and ESO VLT+VIMOS.
It is important to remark,
that luminous Sub-mm source at high z imply in the rest-frame
luminous IR sources. Thus, probably we are studying the same type of objects
in both programs. We have already  observed 3 high z BAL QSOs. \\


\section{GEMMINI OBSERVATIONS AND DATA REDUCTION}
\label{gobser3}

\subsection{Gemini GMOS-IFU observations} \label{ggmosobservations1}

The three-dimensional (3D) deep optical spectroscopy of  the nucleus and the
more extended arc of Mrk\,231 was obtained
during two photometric nights in  April 2005, at the
 8.1 m telescope in Gemini North Observatory.  
The telescope was used with the  Gemini Multi Object Spectrograph (GMOS;
Hook et al. 2002) in the mode integral field unit (IFU; Allington-Smith
et al. 2002).
The spectra cover all the optical wavelength range:
from 3400 \AA to 10000 \AA.
The observations were made in photometric conditions with
seeing in the ranges $\sim$0.4--0.6$''$ (in the observing run of 2005
April 30) and $\sim$0.7--0.8$''$ (in 2005 April 6).
For detail of each observations, see Table 1.

The data were obtained with the IFU in one slit mode, which provide
a spatial field of 3.5$'' \times$ 5.0$''$ for each resulting science data cube.
With this observing configuration, the GMOS IFU is comprised of 750 fibres;
each spans a 0\farcs2 hexagonal region of the sky.
Five hundred fibres make up the 3.5$'' \times$ 5.0$''$  science field of view;
and 250 fibres make up a smaller, dedicated sky field, which is fixed at 1$'$
of distance of the science position (Allington-Smith et al. 2002).
We used in the first night the R831 grating in GMOS, which has a
$\sim$40 km s$^{-1}$ spectral resolution.
In addition, the B600 grating was used in the second night, which has
a $\sim$120 km s$^{-1}$ spectral resolution.

Very deep 3D spectra were obtained for all the observations the B600 grating,
for this bright QSO. The typical
exposure time were of $\sim$1 hs. (for the nucleus and for the arc, see
for details Table 1). These very deep observations were performed mainly
in order to study: multiple components in the OF/BAL process, the spectrum
of the host galaxy, the extreme nuclear starburst (with massive
star population) and the knots in the expanding shells/bubbles.

\subsection{Reduction of Gemini GMOS--IFU data} \label{reductions2}

The following software packages were used to reduce and to analyse the
GMOS-IFU data:
{\sc R3D + EURO3D} visualization tool\footnote{{\sc R3D} is the imaging
analysis software facility developed by us at Calar Alto Observatory
(Sanchez \& Cardiel 2005; Sanchez 2006).
{\sc EURO3D} visualization tool is a software package
for integral field spectroscopy, developed by EURO3D Research Training
Network (Sanchez 2004)};
{\sc IRAF}\footnote{{\sc IRAF} is the imaging analysis software 
developed by NOAO}; 
and {\sc GEMINI}\footnote{{\sc GEMINI} is the reduction
and analysis software facility developed by Gemini Observatory}.

The 3D GMOS-IFU spectroscopic observations were reduced using mainly a modified
version of R3D software package (Sanchez \& Cardiel 2005; Sanchez 2006).
This reduction process was performed following the standard steps:
(1) the data were bias subtracted;
(2) the expected location of the spectra were traced on a continuum lamp
exposure obtained before each target exposure;
(3) the fiber-to-fiber  response at each wavelength was determined from a
continuum lamp exposure;
(4) wavelength calibration was performed using arc lamp spectra and the
telluric emission line in the science data;
(5) the sky background spectrum was estimated before subtraction by
averaging spectra of object free areas;
(6) the calibration flux was done using the observation of standard stars; and
(7) the observations of the nucleus and the arc (for the corresponding gratings)
were then combined in mosaics.
A total of 14700 spectrum --of Mrk 231 and sky-- were reduced and calibrated,
using this technique.

After this basic reduction process, a data cube was created for each
exposure and for each mosaic. The cubes were then recentred spatially by
determining the position of the very nucleus of Mrk 231. This recentering
corrects for differential atmospheric refraction. It is important
to note, that we already performed a detailed study
of the effect of  the atmospheric refraction  in the position of the
very nucleus in 3D spectroscopy (for different wavelengths: from 7800\AA,
to 4200\AA; Arribas et al. 1997, their Fig. 4a). From this study, a
variation in the position of the very nucleus of $\sim$1.5$''$ was found,
at La Palma WHT. Therefore, for the very high spatial resolution GMOS data
(and covering a very large wavelength range) it is important the recentering
process.

To generate two-dimensional maps of any spectral feature (intensity,
velocity, width, etc.) the IDA and INTEGRAL software tools
(Garc\'{\i}a-Lorenzo, Acosta-Pulido, \& Megias-Fernandez 2002) were used.
We have found that the IDA package gets better
results recovering 2D maps from low signal-to-noise data. The IDA
interpolation is performed using the IDL standard routine TRIGRID, which
uses a method of bivariate interpolation and smooth surface fitting for
irregularly distributed data points (Akima 1978).
Maps generated in this way are presented in the following Sections.

The emission line components were measured and
decomposed using Gaussian profiles by means of a non-linear
least-squares algorithm described in Bevington (1969). In
particular, we used the software {\sc SPECFIT}\footnote{{\sc SPECFIT} was
developed and is kindly provided by Gerard A. Kriss.}, and SPLOT from the
{\sc STSDAS}\footnote{{\sc STSDAS} is the reduction and analysis software
facility developed by the Space Telescope Science Institute.}, and IRAF
packages, respectively.
An example of SPECFIT deblending, using three components for each emission
line in IRAS\,01003$-$2238, was shown in figure 2 of L\'{\i}pari et al.\
(2003). We note that in each GMOS spectrum the presence of OF components and
multiple emission line systems were confirmed by detecting these systems
 in at least two or three different
emission lines ([N {\sc ii}]$\lambda$6583, H$\alpha$, [N {\sc ii}]\
$\lambda$6548, [S {\sc ii}]$\lambda\lambda$6717/31,
[O {\sc i}]\ $\lambda$6300, [O {\sc iii}]\ $\lambda$5007, and H$\beta$).
For the study of the kinematics, the {\sc ADHOC}\footnote{{\sc ADHOC} is
a 2D/3D kinematics analysis software developed by Marseille Observatory.}
software package was also used. \\

\subsection{The GMOS IFU Mosaic of Mrk 231}
\label{3.7}

Fig. 1 --NOT V wide field image-- shows together  the whole merger and
the observed GMOS mosaics (in orange colour, and  covering an area of
$\sim$3\farcs5$\times$9\farcs0, $\sim$3 kpc $\times$ 7.3 kpc).
This mosaic was constructed by combining  two individual Gemini +
GMOS-IFU frame (of 3\farcs5$\times$5\farcs0).
The first GMOS frame was centered in the very nucleus of Mrk 231,
and the second one in the more extended bubble/arc
(at 4$''$ to the south, from the very nucleus).

Fig. 1 shows that Mrk 231 consist of a nearly elliptical main body
(of R $\sim$ 10 kpc)
with a compact nucleus, plus two faint tidal tails (see for details
Hamilton \& Keel 1987; Neff \& Ulvestad 1988; Lipari et al. 1994).
Therefore, the GMOS mosic cover mainly the nuclear and circumnuclear region
of the IR merger.

\subsection{Multiple components in the GMOS-IFU emission lines spectra}
\label{res-3.8}

An important point in the study of spectra obtained with relatively high
spectral resolution is the analysis of multiple components (in each emission
line, especially in the stronger lines [N {\sc ii}]$\lambda$6583 and H$\alpha$).
This type of detailed study was performed for several nearby galaxies,
including systems with out flow process, e.g.
NGC  1052, 3079, 3256, 4550, 5514, 7332,
Cen A,  etc (e.g. Plana \& Boulesteix 1996; L\'{\i}pari et al. 2000,
2004d, 2005a; Veilleux et al. 1994; Bland, Taylor, \& Atherton 1987).
For Mrk 231, using the 3D GMOS spectra with high and moderate  spectral
resolution (R831 and B600 grating with spectral resolution of $\sim$40 and
120 km s$^{-1}$, respectively ),
a detailed study of multiple emission line components was
performed (especially in order to analyse OF motion). Fig. 2 
show the presence of these multiple emission (and stellar absorption)
line systems.

In particular, from this study --of multiple component-- the following
main results were obtained:

\begin{enumerate}

\item
{\it  Main Component in the Emission Lines (MC-EMI, MC-EMI$^*$)}:

In the nucleus and the circumnuclear region of Mrk 231 mainly a single
strong/main emission line component (MC-EMI) was detected.
This ELC was measured and deblended using the software SPLOT (see previous
section 3.2).
For this MC-EMI a redhift at Z $=$ 0.04250 (12750 km s$^{-1}$) was
measured.

In addition,
at the south-west nuclear and circumnuclear regions (of the GMOS mosaic)
this main component (MC-EMI$^*$) was detected with a clear blueshift, at 
Z $=$ 0.04220 (12600 km s$^{-1}$).

\item
{\it Blue OF Emission Components (OF-EB1, OF-EB2$^*$, OF-EB3$^*$}:

The presence of several strong OF components were observed,
specially in the main knots of the super giant shells.  These ELCs are
blueshifted, in relation to the systemic velocity and the MC-EMI
of the merger; and they  were deblended using the software SPECFIT.

\begin{itemize}

\item
{\it OF-EB1}:
This is a low velocity blue OF component, which was detected
in almost all the main knots of the 4 super giant bubbles.
We have measured  for OF-EB1 a range of values of redhifts
Z $=$ [0.04140, 0.04210] (12470 , 12630 km s$^{-1}$),
$\Delta$V $=$ V(OF-EB1) - V(MC-EMI) $=$  [--150, --300] km s$^{-1}$.

\item
{\it OF-EB2$^*$}:
This is an intermediate velocity blue OF component, which was detected
only in the main knots located in the south west region of the 4
more external super giant bubbles.
We have measured  for OF-EB2$^*$ a redhift Z $=$ 0.04080 (12240 km s$^{-1}$),
$\Delta$V $=$ V(OF-EB2$^*$) - V(MC-EMI$^*$)  $=$ --400 km s$^{-1}$.

\item
{\it OF-EB3$^*$}:
This is an extreme blue OF component, which was detected mainly
in the south west border of our GMOS mosaic (we call this region as
SW1, which is located at [1.7$''$ West, 5.6$''$ South]).
We have measured  for OF-EB3$^*$ a redhift Z $=$ 0.03920
(cz $=$ 11774 km s$^{-1}$),
$\Delta$V $=$ V(OF-EB3$^*$) - V(MC-EMI$^*$) $=$ --905 km s$^{-1}$.

\end{itemize}

\item
{\it Reed Outflow Component in the Emission Line (OF-ER1)}:

This is a low velocity red OF component, which was detected mainly
in the circumnuclear areas and in some knots of the supergiant bubbles.
We have measured  for OF-ER1 a range of redhifts
Z $=$ [0.04298, 0.04333] (12895 , 12998 km s$^{-1}$),
$\Delta$V $=$ V(OF-ER1) - V(MC-EMI) $=$ [+150, +250] km s$^{-1}$.

\end{enumerate}

In conclusion, with the spectral resolution of this study we can identify
at least 6 different emission line  systems. These results are specially
important for the study of the OF process of the multiple
expanding supergiant bubbles/shells, and also for the generation and
interpretation of the velocity fields and emission line ratios maps.


\section{HST-ARCHIVE AND OUR PREVIOUS OBSERVATIONS}
\label{paobser4}

It is important to remark that the high spectral and spatial resolution GMOS
observations --of Mrk 231-- were obtained mainly in order to continue and to
test our previous studies (which were performed mainly at moderate/low
spatian and spectral resolution, but for wide spatial fields).
Therefore, in this work is essential to compare the Gemini results
with those obtained previously.

The technical details of our previous observations of Mrk 231 (obtained at
La Palma/WHT,  La Palma/NOT, and Keat Peak/Gold-Cam) and the used archive
data (from HST/WFPC2, HST/ACS, HST/NICMOS, and HST/FOS) were already
descrived by Lipari et al. (2005a). However, in Table 1 a summary of
these observations are presented.
In addition, it is important to note that
Lipari \& Terlevich (2006) presented an evolutionary model
for AGNs and QSOs, in which a main point is the composite nature in the
very nucleus of BAL + IR + Fe II QSOs. Thus, the Gemini plus previous
data of Mrk 231 are an important test for the evolutionary
theoretical model of composite AGNs/QSOs.

\clearpage

\section{Decoupling the GMOS-IFU nuclear spectrum: BAL, host and QSO components}
\label{results-DecouplingModelo5}

Very recently, a new method for decoupling the spectra of the QSO/AGN from
the host galaxy --using 3D spectroscopy-- was developed by us
(see for references and details Sanchez et al.  2006a,b, 2004; Jahnke et
al. 2004; Wisotzki et al. 2004; Garcia Lorenzo et al. 2005).
Using this technique the clean 3D spectra of the host
stellar population could be obtained.

The pure 3D spectra of the host galaxy combined with stellar population models
will allow us  to analyse the properties of  the nuclear stellar component.
Furthermore,
the presence of a young/blue stellar component --in the very nucleus of
Mrk 231-- was already suggested, in the framework of
extreme galacti wind scenario with multiple expanding shells for this IR +
BAL merger (and for similar BAL + IR + Fe {\sc ii} QSOs; Lipari et al. 1994;
2003; 2005a).

\subsection{Description of the GALFIT 3D Model,  and
the application of this technique to Mrk 231}
\label{results-modelo}

Different techniques have been developed to decouple the main
components of an image. A commonly used method is to fit the
image with 2D models, including template for each different components.
This type of technique is implemented in GALFIT (Peng et al. 2002), a
program for modelling several components in images.
On the other hand,
Jahnke (2002) developed a method for decoupling the host and
nuclear spectra of galaxies, for 1D long-slit spectroscopy.

Integral field (or 3D) spectroscopy combine characteristics of images and
spectroscopy techniques. Thus,
a natural extension of the modelling 2D images (plus decoupling 1D spectra)
is: to split the 3D spectra
data cube in a set of narrow--band images of the width of the spectral pixel
and treat them as individual images.
This technique has been used successfully for the deblending of
QSOs/AGNs with 3D spectroscopy (see Sanchez et al. 2004, 2006a,b; Jahnke et
al. 2004; Wisotzki et al. 2003, 2004; Garcia Lorenzo et al. 2005).

For the nuclear region of Mrk 231 (r $<$ 1\farcs7),
the 3D image modelling of the nucleus and the host galaxy
 for each monochromatic
image was performed using GALFIT 3D. The 3D model comprise a narrow Gaussian
function (to model the nucleus) and a de Vaucouleurs law (to model the
galaxy), both convolved with a PSF. Thus, a main step in this technique is
to obtain the best PSF, for the 3D data (specially, for spectra
obtained with high spatial and spectral resolution).

For the GMOS data, the PSF was carefully obtained from the very nucleus
of Mrk 231, using the H$\alpha$ and H$\beta$ broad line emission.
Thus, we are using a PSF, which was derived from the same 3D spectra
that we are studying (similar to a PSF observed simultaneously with the
scientific data: i.e., probably the best PSF available).
In particular, the PSF was generated using the technique described in detail
by Jahnke et al. (2004).  For Mrk 231, the H$\alpha$ and H$\beta$ have
a mix of broad line and continuum emission. Adding up the corresponding
image slice and subtracting the appropriate background frame removes
all of the contained host and nuclear continuum emission; resulting
in a pure BLR--PSF.

Then, the fitting process was performed twice. A first iteration
were all the morphological parameter of the host galaxy fitted freely,
and second one were they are fix to the average values along the
wavelength (as described in Sanchez et al. 2006).
This method ensure a clean decoupling of both spectra.

In addition, it is important to remark two main point about the
process of modelling the 3D GMOS spectra of Mrk 231 nucleus,

\begin{enumerate}

\item
Only with the set of very deep Gemini GMOS IFU
observations (obtained with B600 grating, see Table 1),
this method --for decoupling the 3D spectra-- allowed a detailed study of 
the  spectral feature of the host galaxy.
Since the study of the faint host galaxy component require 3D spectra with
very high S/N.

Using the GMOS--IFU observations obtained with relatively short exposure
time (i.e., those observed with R831 grating), we found that the pure
spectrum of the host galaxy has very low S/N. Even for exposure
of 900s in a 8m class telescope (and for the nucleus).

\item
For the GMOS data obtained with B600 grating,
we found that in almost all the observed spectral range
the derived spectrum of the host galaxy shows high S/N ratios (larger than 10).
Even at $\sim$3700 \AA (close to the limit of blue wavelength range),
the spectrum of the host galaxy shows good S/N (larger than 5).
Obviously this fact is due to the ``very" large exposure time of the B600
3D spectra (with more than 1 hs. of exposure time, for the nucleus).

On the other hand, it is important to remark that it is already known
that the nuclear/QSO spectrum of Mrk 231 shows a strong continuum fall at
wavelength shorter than 3800 \AA (see Fig. 3, where the arrow depicts
the strong continuum fall at this wavelength).
Therefore, the blue limit for our 3D decoupling method was in the range
$\lambda\lambda$3700--3800 \AA.

\end{enumerate}

\subsection{Decoupling the BAL, Host Galaxy and QSO components in Mrk 231
(using GALFIT 3D Model)}
\label{results-modeloc}

Figures 4 and 5 show the spectra of the QSO and the host galaxy of
Mrk 231 for the nuclear region (r $\sim$ 1\farcs7). Which were obtained
using the technique of decoupling described in the previous sub-Section
and from the 3D GMOS+B600 data.
From this study and these plots, the following main results were found:

\begin{enumerate}

\item
{\it For the QSO component:}
Figs. 5a, b, c clearly show --specially at H$\alpha$-- that ``mainly" the broad
emission lines are present, and thus the standard narrow line region (NLR) is
absent.
This is a very interesting property found in BAL IR QSOs
 (see for reference  Veron et al. 2006; Lipari \& Terlevich 2006;
 Turnsheck et al. 1997; Lipari 1994). \\

\item
{\it For the host galaxy component:}

\begin{itemize}

\item
Fig. 5c depicts that the host galaxy has a strong nuclear blue/starburst
component. Specifically, the flux of the host galaxy shows a very
strong increase in the blue region of the optical spectrum.
In addition, a similar result was found from the study of the
emission and absorption lines in the individual GMOS spectra
of the nuclear region (see the next Sections).

It is interesting to note that in the blue wavelengths, the pure
QSO spectrum shows a strong fall in the continuum (which is typical of
luminous IR galaxies, and it is associated mainly with strong nuclear
reddening, by dust).

\item
Figs. 5b and c show  that the BAL system {\sc i}
(at Na ID, He {\sc i} and Ca {\sc ii} lines) are clearly
observed in the host galaxy, i.e.: the BAL--I shows an extended
morphology.

\item
Fig. 5a shows --at relatively low S/N-- OF components mainly in
[N II]$\lambda\lambda$6548 and 6583 emission lines, of the host galaxy.

\end{itemize}

\end{enumerate}

An interesting explanation for the point (i)  is that the
multiple nuclear explosive events (composite OF, detected previously 
in this QSO) probably expel the standard NLR.
However, in the next Sections we will present evidence of
narrow emission line systems --in the nuclear and circumnuclear regions--
associated mainly with the OF process.  \\

\clearpage

\section{Mapping with GMOS-IFU and the nuclear
continuum flux and  the extreme blue component}
\label{results-9blue}

In order to confirm one of the main result obtained in the previous
section: i.e., the presence of an extreme blue component
detected in the clean spectra of the host galaxy (for the nuclear region of
Mrk 231), we have analysed
--at different wavelength ranges-- the shape of the continuum emission
in each individual spectrum, for all the nuclear and circumnuclear regions.

Furthermore, it is important to remark that
we already observed --using low spatial resolution 3D La Palma
WHT+Integral spectroscopy-- some variations in the nuclear continuum shape
(even in the red wavelength range, adjacent to H$\alpha$).
Specifically,  clear differences in the continuum shape were observed
among the Fig. 6a, b, c and Fig.  6e of Lipari et al. (2005a), at different
nuclear and circumnuclear areas.

Thus, a first basic  qualitative study of GMOS spectra was performed, which
was based in a direct/simple inspection of the continuum shape, at each
spectrum.
Figures 6 and 7 show two sequence of individual spectra
(for the Blue + [O {\sc ii}]3727 and Visual + H$\beta$ wavelength regions)
along the north-south direction  (PA $=$ 00$^{\circ}$).
From this qualitative study of all the nuclear spectra, two very interesting
results were found (which are evident in Figs. 6 and 7):
(i) at the  southern nuclear area (from the very nucleus)
an extreme blue continuum component was detected;
(ii) at  the  northern area an extreme red continuum component
was observed.
This results were verified at almost all the observed wavelength ranges.

Then,
a detailed quantitative study of the continuum was performed,
using for this purpose a colour index defined --by us--  as
the difference of fluxes at the border of the wavelength range
of each GMOS CCD (using the B600 grating; see Table 1 for details
of the GMOS observation,  and Allington-Smith et al. 2002 for
details of the GMOS instrument),

\begin{itemize}

\item
For the Blue Wavelengths:

\centerline{[Flux($\lambda$4300) -- Flux($\lambda$3700)] $\times$ 10$^{16}$,}

\item
For the Visual Wavelengths:

\centerline{[Flux($\lambda$5250) -- Flux($\lambda$4350)] $\times$ 10$^{16}$.}

\end{itemize}

Fig. 8 shows a map of the continuum colour index, for the wavelength
region around H$\beta$ (i.e., using the second colour index).
This GMOS map  confirm the presence of a extreme
blue component in the continuum flux, which is located to the
south of the nuclear region. This area is coincident with the region where
we previously suggested that the galactic wind --with super bubble/shells--
is cleaning the nuclear dust,
and thus this fact is probably allowing to see the extreme nuclear starburst.

The contour of this continuum colour index map (Fig. 8b) shows
a clear peak in this blue continuum component, which is located at 0\farcs3 to
the south of the very nucleus. This position of the peak is coincident with
the location of the more internal ring S5 (with
also a radius of 0\farcs3; Lipari et al. 2005a).
Thus, this peak is probably positioned inside of the dusty ring or shell 
 S5. Furthermore,
there is a symmetric red continuum peaks, positioned
at 0\farcs3 to  the north of the very nucleus.
The individual GMOS spectra of these two symmetric blue and red peaks show
strong and narrow multiple emission line components
(with OF), specially in the lines:

(i) [O {\sc ii}]$\lambda$3727  (Fig. 6): the NLR in this line is absent
at the very nucleus, but it is very strong and with double components
at  [0\farcs4 south, 0\farcs2 east], [0\farcs4 south, 0\farcs0] from the
very nucleus (which are in  the area of the strong blue continuum peak).
We have measured for these two [O {\sc ii}] component a
$\Delta V =$ --400 km s$^{-1}$, and FWHM of 160 km s$^{-1}$.

(ii) IR Ca {\sc ii}$\lambda$8500 triplet: these lines  show strong
relatively narrow stellar emission and absorption   ``only"
at  0\farcs4 to the north and south of the very nucleus (the areas of the red
and blue continuum peaks).
In the next Section, this point (the emission and absorption of
the IR Ca {\sc ii} triplet) will be analysed in detail.

(iii) [S {\sc ii}], [O {\sc i}]6300, H$\alpha, $H$\beta$, [O {\sc iii}]5007:
in Section 10, a detailed study of the NLR
-at these lines- will be presented, for the very nucleus and the nuclear
region. Mainly,
we found that the nuclear NLR is associated to the nuclear OF process. \\

These results --specially the points (i) and (ii)-- are the clear signature
of a very young stellar population (close to the very nucleus).
More specifically,
the presence of two strong narrow [O {\sc ii}]$\lambda$3727 components
(plus the emission line ratios found in this area) could be associated
 mainly with young H {\sc ii} regions
 with strong OF process (Lipari et al. 2000, 2004a,d).
The strong narrow emission plus absorption in the IR Ca
{\sc ii}$\lambda$8500 triplet is clearly associated with the peak of 
red super-giant (RSGs) activity, with age of: 8 Myr $<$ age
$<$ 15 Myr (see for references Lipari \& Terlevich 2006).

On the other hand, different previous studies performed with
very high spatial resolution at radio and millimetre wavelengths
(using mainly interferometric techniques)
already proposed that in this area --of the ring S5-- there is a 
disk of molecular gas, with extreme star formation (SFR) rate of
$\sim$100--200 M$_{\odot}$ yr$^{-1}$ (Bryan \& Scoville 1996;
Downes \& Solomon 1998; Carrilli et al. 1998; Taylor et al. 1999).
Thus, our finding of a dusty ring or toroid of extreme and young star
formation process around the nucleus of Mrk 231 is in excellent agreement
with previous studies performed at different wavelength regions and
using different observational techniques. \\

\clearpage

\section{Mapping with GMOS-IFU the nuclear emission of the near-IR
Ca {\sc ii} triplet}
\label{results-irca2tri}

M231 is one of the few QSOs that show very strong near IR
broad Ca {\sc ii} triplet in emission (see for details and references
Lipari \& Terlevich 2006). Furthermore, our evolutionary and
composite model for QSOs/AGNs predict the simultaneus ocurrence
of strong near-IR Ca {\sc ii} broad emission and strong Ca {\sc ii}
stellar absorption, in young Fe {\sc ii} QSOs. Thus, the study of
the high resolution near IR GMOS data
(of Mrk 231) is: (i) an important test of our composite and evolutionary
model of BAL + IR + Fe {\sc ii} QSOs; (ii) an important tool in order to
study in detail the physical conditions in the nuclear and circumnuclear
regions.

It is important to remark, that the GMOS IFU spectra of the near-IR
Ca {\sc ii} triplet were obtained with the best spectral resolution
of this instrument  ($\sim$40 km s$^{-1}$, using the GMOS grating R831).
In addition, these data were obtained in our observing run with the
best seeing/spatial resolution (see Table 1).
Figure 9 shows the more interesting GMOS data, at the near--IR
Ca {\sc ii} triplet + O {\sc i}, which were selected
from the spectra of all the nuclear region.
From this plot and study, we found very interesting results:

\begin{enumerate}

\item
In the very nucleus of Mrk 231: mainly a very broad blend of strong emission
in the IR Ca {\sc ii} triplet + O {\sc i} was detected
 (Fig. 9).

\item
At 0\farcs4 north and 0\farcs2 west, from the very nucleus of Mrk 231:
strong  narrow  emission in the IR Ca {\sc ii} triplet
+ O {\sc i} superposed with the broad blend -of Ca {\sc ii} emission-
were clearly detected (Fig. 9).
The spectra of this region show the typical feature of a ``Seyfert 1.5"
AGN.

For the  $\lambda$8446 O {\sc i} and the IR $\lambda$8498, 8542 Ca {\sc ii} lines,
the following values of FWHM were measured: 190 and 370 km s$^{-1}$, respectively.
It is important to note that in the GMOS R831 spectra (Fig. 9)
the  $\lambda$8662 Ca {\sc ii} line  is positioned in the
wavelength area of strong near IR sky lines. Thus, this Ca {\sc ii}8662 line
(superposed with strong sky lines) was not used for this study.

\item
At 0\farcs2 north, from the very nucleus of Mrk 231:
strong relatively narrow  absorption superposed to the narrow emission
in the IR Ca {\sc ii} triplet were clearly observed (Fig. 9);
specially, in the Ca {\sc ii} $\lambda$8498 line. 

It is interesting to remark that Fig. 9 shows in the Ca {\sc ii} triplet
the absorption features already suggested by Lipari \& Terlevich (2006),
in their evolutionary unification composite model: i.e., the superposition
of the absorption and emission line; which is more clear at
the line Ca {\sc ii} $\lambda$8498 (as an {\bf ``unusual absorbed -or double-
peak"}).
The reason for this unusual observed feature --at Ca {\sc ii} $\lambda$8498
line-- is the interesting fact that the emission fluxes in the triplet is
almost 1:1:1, but the absorption is 1:9:5. Thus, in this
Ca {\sc ii} $\lambda$8498 line the emission and the absorption could be
detected (together) more easily.
In addition, for the same reason the line Ca {\sc ii} $\lambda$8542 is
the more absorbed (Ca {\sc ii} line). This feature is also clearly observed in
Fig. 9 where the line $\lambda$8542 is almost absent.

It is important to note that for the very near area, at 0\farcs4 north
and 0\farcs2 west (from the very nucleus) the peaks of the Ca {\sc ii}
$\lambda$8498 and 8542 lines do not show any unusual double peak/feature.

\item
At 0\farcs4 south, from the very nucleus of Mrk 231:
also strong  narrow  absorption superposed to the narrow emission
in the IR Ca {\sc ii} triplet were observed (Fig. 9).
In particular, the superposition
of the absorption and emission line is more clear at
the line Ca {\sc ii} $\lambda$8498 (as an unusual  ``double peak").

\end{enumerate}

It is important to remark, that these interesting areas
at 0\farcs4 to the North and South of the very nucleus (i.e., the
{\bf only regions} where we detected
the IR Ca {\sc ii} triplet with ``Seyfert 1.5 features
and narrow ``absorption plus emission" lines) 
could be associated with the
previously detected ring S5 (Lipari et al. 2005a).
In this ring, the GMOS data also show:

\begin{enumerate}

\item
Two very strong peaks of blue and red continuum components
(located at symmetric position, $\sim$0\farcs3 from the very nucleus).

\item
Very strong multi emission line components (with an
OF of --400 km s$^{-1}$) in the [O {\sc ii}]$\lambda$3727 lines,
and at the position of the blue continuum peak.

\end{enumerate}

These finding are the typical evidence or signature of very young stellar
population (or nuclear starburst), probably
in a dusty SB shell or ring/toroid.
More specifically:
the strong  narrow {\it ``emission plus absorption" in the IR Ca
II$\lambda$8500 triplet} is clearly associated with the peak of 
RSGs activity, with age of,
{\bf 8 Myr $<$ age $<$ 15-20 Myr}, in metal-rich stellar populations (see
for details Terlevich, Diaz \& Terlevich 1991; Lipari \& Terlevich 2006).
This fact is an interesting prediction of the Evolutionary and Composite
Model for QSOs/AGNs (Lipari \& Terlevich 2006), which was confirmed using
very deep high resolution GMOS 3D spectroscopy.

On the other hand,
the presence of a red peak of the continuum with narrow stellar emission
plus absorption in the IR Ca {\sc ii} triplet lines
(and thus associated with a very young stellar
population) could be explained with the presence of large
amount of dust in this type of very young SB area, of the shell/ring
(Lipari \& Terlevich 2006). \\

\clearpage

\section{Mapping with GMOS-IFU the BAL  {\sc i} and {\sc ii} systems}
\label{results-BAL6}

We have already started a study of the BAL systems, in Mrk 231, using 3D
and 1D spectroscopy with moderate spatial and spectral resolution
($\sim$1.0--1.5$''$ and $\sim$100 km/s, respectively;
Lipari et al. 2005a). More specifically, we have studied: (i) the H$\alpha$
blue emission bump, which is at the same OF velocity of the BAL {\sc i}
systems, and probably associated with this BAL {\sc i};
and (ii) the light curve variability of the BAL {\sc iii} system, probably
associated with SN or HyN.

In the present work,
we have continued --with GMOS-- our previous study of the BAL systems of
Mrk 231, but using: the optical {\bf absorption} lines, and improving the
spectral and spatial resolution.
We note that very deep Gemini GMOS-IFU/B600 data were obtained in the
observing run of 2005 April 30, i.e., with also the best spatial resolution
of our observations (with the best seeing of
$\sim$ 0\farcs4--0\farcs5, FWHM). Mainly, these very deep GMOS data
(with high S/N) were used for the study of the BAL systems of Mrk 231.

\subsection{Study of individual 3D GMOS--IFU spectra}
\label{results-BAL7}

We have studied first the individual 3D Gemini GMOS spectra,
for the following reason: in the  region of the line
Na ID$\lambda$5889-95 we can see at the same time (without blending)
the Na ID BAL absorption line and the broad line emission.
This broad Na ID emission is originated in an almost point like BLR,
and therefore it is one of the best trace --or the best upper limit--
for the final instrumental seeing.
Furthermore, the line Na ID$\lambda$5889-95 shows the
strongest absorption of all the BAL observed in Mrk 231. Thus,
using very deep 3D Gemini GMOS spectra, for the nearest BAL QSO, and
for the strongest absorption line (Na ID BAL I): the high S/N
of these 3D spectra allowed a detailed study of the extension of the BAL I
system.

Fig. 10  shows a sequence of individual 3D Gemini GMOS IFU spectra,
along the north-south direction, from the very nucleus (at the position
angle PA $=$ 00$^{\circ}$),
 with a step of 0\farcs2. This plot clearly shows that the Na
{\sc i}D$\lambda$5889-95 BAL system {\sc i} is extended: reaching
a radius of $\sim$1.4--1.6$''$ ($\sim$1.2--1.3 kpc).
This plot also shows that  the broad emission of Na ID only reach
$\sim$0.6--0.8$''$ (see Fig. 10).
A similar behaviour was found, for the Na ID BAL system I, at several
position angle (PA) including  the PA in the direction of the jet
(PA$=$--120$^{\circ}$; Ulvestad et al. 1999a,b).

In addition, it is important to remark that Fig. 10 also shows the presence
of narrow and weak Na ID absorption line at the redshift of the host
galaxy (at the same wavelength position of the broad Na ID emission).

At the blue wavelength regions, we have verified that the BAL system {\sc i} shows
the same extended nature, in the absorption lines 
Ca {\sc ii} H$\lambda$3969, Ca {\sc ii} K$\lambda$3933, and He I$\lambda$3889.
In the next Section a detailed study of the maps of some of these
absorption lines will be presented.

\subsection{GMOS-IFU absorption maps for the BAL {\sc i} and {\sc ii} systems}
\label{results-BAL7.2}

Figure 11(a), (b), (c), (d), (e) and (f) show  the GMOS maps and contours
for the absorption of the BAL {\sc i} and {\sc ii} systems. These maps were constructed
with the  technique described in Section 3: i.e. we have measured for the
BAL {\sc i} and {\sc ii} systems the absorptions at each spectra, then with a
Table of absorption fluxes and the positions (of each spectra) these data were
converted to FITS maps using the software  IDA (see for details and
references Section 3).

It is important to note, that first we fit
the absorption --in each spectra- using the main component.
We have verified that this way is a simple and efficient way
to detect  small elongation. Since the study of absorption
with multiple component, could include some problems (for example
the selection of the number of components, etc); which could be even more
important than the very weak elongation (that we are searching).

In particular, these figures present the following interesting results:

\begin{enumerate}

\item
Fig. 11a, b depicts the map for the BAL {\sc i} at the line Ca {\sc ii}
K$\lambda$3933.
These two plots clearly show (specially the contour map) that the absorption
of the BAL {\sc i} system is elongate at the PA close to the radio jet
direction (at PA $\sim$ -120$^{\circ}$).

\item
Fig. 11c, d shows the map for the BAL {\sc ii} at the line Na ID. 
Again, these two plots clearly depict that the absorption
of the BAL {\sc ii} system is elongate at the PA close to the radio jet direction
(at PA $\sim$ -120$^{\circ}$).

\item
Fig. 11e, f depicts the map for the BAL {\sc i} at the line Na ID. 
In this case is evident that the fit (of the absorption lines)
with only one component did not allow to detect any elongation.
The main reason for this fact is that the GMOS B600 and R831
spectra of this ``very strong" BAL {\sc i} system (at Na ID), required
for the fit more than one component. However we already noted that
the use of multiple components for the fit of this absorption
present several problems, since the small elongation (that
we are searching) could be masked by the errors in the technique
of deblending.

\end{enumerate}

Therefore, the results obtained in this Section are in excellent agreement
with the previous study of H$\alpha$
emission (using 3D spectra with moderate spatial resolution; Lipari et al.
2005a), where we found an OF emission bump, with a peak at the same velocity
of the BAL system {\sc i} (--4700 km s$^{-1}$).
This H$\alpha$ emission bump was detected in the nucleus and at 0.6--1.5$''$
(from the nucleus) in the direction of the radio jet (at PA $=$
--120$^{\circ}$ and 60$^{\circ}$, see their Table 4).
Thus,
these results are consistent with the fact that the blue H$\alpha$ bump
and the BAL--I system could be associated with OF clouds of the small
scale jet.
A detailed discussion of this point will be presented in Section 13.2.

\subsection{Study of the BAL systems with GMOS high spectral resolution}
\label{results-7BALR831}

Figure 12a shows the BAL {\sc i} and {\sc ii} systems in the strong line Na ID,
for the best spectral resolution of GMOS-IFU: 40 km s$^{-1}$ (using
the grating R831). From this plot, we found interesting
results:

\begin{enumerate}

\item
The BAL systems {\sc i} and {\sc ii} do not show  strong/clear variability
in the last 3 decades, since all the spectra obtained with similar high
resolution (e.g., Rudy, Foltz \& Stocke 1985; Kollatschy, Dietrich \& Hagen 1992;
Forster, Richt, McCarthy 1995; Rupke et al. 2002) are very similar to
that presented in Fig. 12a.

\item
The BAL {\sc i} system shows clearly the presence of multi absorption components.
Previously, Forster et al. (1995) suggested that the fit of the strong
Na ID absorption line require at least 9 components, for the study of
this BAL {\sc i} system.
Rupke et al. (2002, their Fig. 11) plot a fit of the BAL
system {\sc i} --at the Na ID absorption line-- using 12 components.

\end{enumerate}

It is important to note, that all these previous work fit with only one
component the Na ID BAL {\sc ii} system.
This point is important in order to confirm the result obtained in
the sub-Section 8.2, since we detected the weak elongation in the
maps of the BAL systems, only when we fit the absorptions with only
one component (i.e., when we avoid the problem of fitting the absorption
line with a large number of components).

\subsection{Deep GMOS-IFU observation of the Na {\sc i}D BAL
{\sc iii} systems: confirming the exponential variability/fall}
\label{results-opbal3}

The variability of the short lived BAL {\sc iii} Na {\sc i}D system was
previously studied (by Lipari et al. 2005a).
We found that the BAL {\sc iii} light curve (LC) is clearly asymmetric with:
a steep increase, a clear maximum and an exponential fall
(similar to the shape of a SN LC).
An important point --in the present work-- is to confirm the exponential
fall of this BAL {\sc iii} system, using very deep GMOS data.
Since this exponential fall is a very important fact, in the explosive
scenario of Mrk 231.

The optical Na {\sc i}D BAL systems of Mrk 231 was observed
using 3 different GMOS grating configuration (see for details
Table 1). Only with very deep GMOS spectra and using the grating
B600 (with good transmission and spectral resolution, at the
region of the Na {\sc I}D BALs) we detected this very weak
BAL  {\sc iii}  system: in 2005 April.

Fig. 12b depicts this weak Na {\sc I}D BAL  {\sc iii}  system.
In order to confirm the behaviour of the variability of this  absorption
system,  we have measured the equivalent width ratio  of
BAL {\sc iii}/(BAL {\sc i} + {\sc ii}).
For this ratio a value of 0.007 was obtained, for 2005 April 30.
This new result was combined with those obtained previously by
Lipari et al. (2005a; their Table 5) in 
Fig. 12c. This plot  shows the shape of the BAL {\sc iii} light curve fall
(including the new GMOS/2005 observation), plus the best fit.
Again, we obtained the  best fit of the LC fall, using
an exponential function.

Thus, this study (using deep GMOS data) expand and confirm the previous
study of variability of the Na ID BAL {\sc iii} system
(covering almost all the period in which this system appeared).
In general, the exponential shape of the BAL {\sc iii} system LC
is similar to those LC of  SNe flux, emission line, etc. \\

\clearpage

\section{GMOS-IFU data of the nuclear expanding supergiant shells}
\label{results-1map9}

Mrk 231 is probably the proto--type of exploding QSO. One of the
main evidence of this explosive process is the presence of 4
nuclear expanding supergiant shells. In this work,
the main knots of these multiple nuclear shells will be analized --for
this distant merger-- using mainly GMOS-IFU + HST data of high spatial
and spectral resolution.

It is important to study in detail the main knots detected in the
multiple supergiant bubble with high resolution 3D spectroscopic data.
Since they are  the best and bright tracers of the expanding super bubbles.
In order to define the location of these main knots, in
Fig. 13(a) we present high resolution HST WFPC2 broad-band images of the
nuclear region of Mrk 231;
obtained in the optical wavelengths through the filter F439W ($\sim B$).
This HST image  shows the main concentric super giant galactic shells.
In order to depict in detail the structure --knots-- of the shells,
in Fig. 13(b) we show the result best example  obtained from
the subtraction of a smooth HST image of the main body of Mrk 231
 (for the filter F439W $\sim$B).

The general properties of the 3 more external shells (S1, S2, and S3) were
studied by Lipari et al. (2005a).
In this paper,  we will study in detail the main knots of these shells
(S1, S2, and S3) plus
the nature of the inner shell S4 and the ring S5 
(using high spatial and spectral resolution GMOS data).

It is important to note that
a possible ring or shell S5 was observed in the  HST optical (I-B)
colour image, mainly as a dusty ring (see Lipari et al. 2005a:
their Fig. 4). Which is located very close to the nucleus, with
a radius $R_{S5}\,= 0\farcs3 = 0.2$\,kpc.
In this paper (sections 6, 7 and 13.1) we are presenting interesting
results about this ring S5.

\subsection{GMOS--IFU maps and cubes of nuclear region}
\label{results-5.2}

Figures 14(a), (b), (c), (d), (e) and (f) show
the red, green continuum and
the  H$\alpha$, H$\beta$,  [O {\sc iii}]$\lambda$5007
and [S {\sc ii}]$\lambda$6717+31
emission line images, obtained from the GMOS cubes.
These figures show strong continuum (adjacent to  H$\alpha$ and H$\beta$,
respectively) and emission line, asociated mainly with the nucleus and
the more external superbubble/arc S1.

From these GMOS images, we remark the following new interesting
feature,

\begin{enumerate}

\item
The  H$\alpha$, H$\beta$, and [S {\sc ii}]$\lambda\lambda$6717 + 31
emission images of the nucleus clearly show an asymmetry or  elongation
to the east and at $\sim$1$''$, from the very nucleus.
The spectra in this region --call by us E1-- show H {\sc ii} region
emission line ratios (see for detail Section 10). Furthermore,
this region E1 is located very close to the position of the
knot K3 (which is the main knot of the shell S3). This knot K3 also
shows H {\sc ii} region emission line ratio.

\item
The external shell S1 shows a large and strong complex or
association of several knots.
This complex is located at 3\farcs3 (3 kpc) to the south-west (of the
nucleus) and the main members of this association are: the blue knots
K11 and K12 and the red one K14 (see Figs. 14 and 13).

\item
The  [O {\sc iii}]$\lambda$5007 emission line images
shows a strong emission, located very close to the knot K14 (at the
north west border of the strong complex of H {\sc ii} regions, in S1).
Specifically,
the position of the  strong [O {\sc iii}] emission is --within
the errors--  almost coincident with the position of this knot K14.
The study of the individual 3D spectra in this region shows
that all the main knots in this area/complex have strong and multiple
blue shifted components, mainly in the [O {\sc iii}]$\lambda$5007 emission
line.

Thus,
this region will be associated with a strong source of outflow, at
the border of the super bubble S1. This OF process will be studied using
the kinematics maps (combining the Gemini+GMOS and La Palma WHT+Integral
 3D spectroscopy data).

\end{enumerate}

In the next Sections, the main kinematics and physical  properties
of the main knots/complexes of H {\sc ii} regions in the super giant bubbles
will be analysed (using individual 3D GMOS spectra and maps). \\

\subsection{GMOS-IFU spectra of the knots in the expanding supergiant shells
(of Mrk 231)}
\label{gspectra}

Table 2 and Fig. 13 present the location and properties  of the strong
knots, which are mainly located in the more external super giant shells S1,
S2 and S3
(we used for these knots the same notation suggested by Surace et al. 1998).
In addition, these knots are labelled in Figs. 13(a) and (b).
Surace et al. (1998) already studied in detail these knots using
HST photometry, in the B and I bands. It is interesting to note, that
recently we started a detailed photometric study of all the HST broad
band images available: U, B, I, and H (Lipari et al. 2007c, in preparation).
The main results of this last study is consistent with those obtained
by Surace et al. (1998).
In this Section,  we will study --for these and new knots-  their GMOS
spectra, and their relation with the OF process in the 4 shells.

Using the GMOS 3D spectra and maps (with a spatial resolution
of $\sim$0.5--0.6$''$, for the observing run of 2005 April 30),
we performed a detailed study of the morphological, kinematics and
physical  conditions of the main knots, of the multiple expanding bubbles.
Figure 15 depicts examples of the red 3D spectra of these
main knots or H {\sc ii} regions/complex.
Tables 3, 4, 5, 6, 7, 8, 9, 10 and 11  show the values of
the fluxes and FWHM  of the emission 
lines and their ratios, for knots in the bubbles.

In order to study the GMOS spectra of the main knots of the supergiant
expanding bubbles we used the following technique: (i) first the main
knots of the 4 more external super giant bubbles were selected, from the high
spatial resolution HST WFPC2 and ACS images; (ii) using the HST position
($\Delta\alpha$ and $\Delta\delta$, off set from the very nucleus) of
the main knots, we then selected the nearest GMOS individual spectrum.
Therefore, there are some differences between the off set obtained by
Surace et al. (1998, using mainly the peaks of the HST WFPC2 B and I images)
and those  presented in Tables 5, 6 and 7. Our off set were derived from the
nearest GMOS spectra (of the corresponding knot peaks).
On the other hand, we have verified also
that the nearest spectrum --corresponding to each knot-- shows the strongest
value of continuum and line emission (for all the area of each knot).

In addition,
it is important to note that only in the very deep GMOS
3D data (with 1 hs. of exposure time, see Table 1) the
spectra depict very high quality, with S/N $>$ 5 in the weak
OF components of the host galaxy. Which is
required in order to study the weak knots of the more internal
super giant bubbles (S2, S3, S4 and S5).
Only in the south-west circumnuclar areas, the OF components show strong
values of flux in the emission lines.
It is also important to remark that in particular in the shell S2, S3 and S4
we have studied several new and interesting knots,
which were included in the Tables of fluxes, with the corresponding off set
--RA and DEC of the nearest GMOS spectra-- from the very nucleus.

From this detailed study of the main knots of the 4 supergiant bubbles of
Mrk 231 (see the Tables 3, 4, 5, 6, 7, 8, 9, 10 and 11), we remark the
following main results:

\begin{enumerate}

\item
{\it General results:}

The 4 external supergiant bubbles show --in $\sim$75 per cent of their
extension-- low velocity OF velocities of $\Delta$V $\sim$ [--150, --250]
km s$^{-1}$, with emission line ratios typical of shocks with
low velocities (Heckman et al. 1990; Dopita \& Southerland
1995).

However, for the south west region (in these 4 bubbles,
with  knots located at almost all the position angles) 
a clear blueshifted  emission was detected, in the main and the OF
components. With a high value of OF, of
$\Delta$V $=$ OF-EB2$^*$ $=$ V(OF-EB2$^*$) - V(MC-EMI$^*$)
$\sim$ --400 km s$^{-1}$.   \\

\item
{\it Supergiant bubble/shell S1: }

For the main knots of this bubble we remark the following results,

\begin{itemize}

\item
Knot 14:
Only in the knot K14-East and K14-West we detected a very strong OF emission
line component, which is similar in flux --or even stronger-- that the
main component (Table 4).  Both components (the OF and MC systems) show
LINER properties associated with shocks.

\item
In the 3 main knots of S1 (K14, K12 and K11) we found high values
of OF velocities, with  $\Delta$V $\sim$ --400 km s$^{-1}$.

\item
At the strong knot K14, clear WR features were detected.
Specially, in the east border of this knot K14 (K14-E, see Table 4).

\item
The other knots  depict mainly LINER properties. In addition,
the  knots K7 shows  composite values of emission
line ratios (ELRs): between LINER and H {\sc ii} regions. \\

\end{itemize}

\item
{\it Supergiant bubble/shell S2: }

For the knots of this shell S2 the following main results were found,

\begin{itemize}

\item
The main/strong knots (of this shell) are all located in the south west
region (knots S2a, S2b, S2c, S2d) with typical emission line ratios of LINER,
associated with shocks of low velocity.

\item
The  knots S2f and S2g show composite values of ELRs,  which are
between LINER and H {\sc ii} regions.  \\

\end{itemize}

\item
{\it Supergiant bubble/shell S3: }

For the main knots of this shell S3 we found the following main results,

\begin{itemize}

\item
The Knot S3a (also called K3, by Surace et al. 1998 and Table 2)
shows values of emission line ratios clearly consistent with
H {\sc ii} regions. This result is
in good agreement with that obtained for the near area E1,
which also shows H {\sc ii} region emission line  ratios.

\item
The other knots (of this shell) show mainly LINER properties,
or composite values between LINER and H {\sc ii} regions. \\

\end{itemize}

\item
{\it Supergiant bubble/shell S4: }

For the main knots of the shell S4, very interesting results were found.
In particular, we remark,

\begin{itemize}

\item
This internal shell show in all the knot two similar emission line
systems with strong [S {\sc ii}]$\lambda$6717/31 and [O {\sc i}]$\lambda$6300.
This are typical feature associated with shocks of low
velocities in a dense medium (similar to those observed in
the OF of SNR and Herving--Haro objects; Heckman et al. 1990;
Binette, Dopita, Tuohy 1985; Canto 1984; Shull \& McKee 1979).

\item
In addition, we have measured for these two emission line systems
an OF with $\Delta$V $\sim$ --200 km s$^{-1}$. Which is consistent
with the presence of low velocity shocks and the strong [S {\sc ii}]
and [O {\sc i}] emission.  \\

\end{itemize}

\end{enumerate}

A detailed study of the emission line ratios of all these shells
and knots will be present in Section 10 (using ELR maps and the
main optical diagnostic diagrams of ionization).


\subsection{Wolf Rayet features in the main complex of knots
in the external shell S1 (GMOS-IFU spectra)}
\label{results-6.5wr}

An interesting result obtained from these spectra (Figure 16
and Table 4) is the detection in the knot K14-East  of
clear Wolf Rayet (WR) features at $\lambda$4650 \AA.
These  WR features suggest the presence of a high number of massive
stars (probably in young super star clusters), with  ages
t $<$ 6--8 $\times$ 10$^6$ yr in this knot (which is associated with
a complex of H {\sc ii} regions: K11 + K12 + K14).
In the next sub-Section, we present evidence that this complex of
H {\sc ii} regions  is probably a point of rupture of the supergiant
bubble S1.

It is interesting to note that
the [N {\sc i}]$\lambda$5199 emission line was also found
in this knot K14-East (where we detected the WR bump).
A similar behaviour (i.e. strong WR and [N {\sc i}] emission)
was observed in the spectra of the prototype WR galaxy NGC 6754
(Osterbrock \& Cohen 1982).

For  a `total'  WR$\lambda$4650 flux of
F$_{WR} =$ 0.39 $\times 10^{-16}$ erg cm$^{-2}$s$^{-1}$ was measured;
and the corresponding luminosity is
L$_{WR} = 1.31 \times 10^{38}$ erg s$^{-1}$. This last value is 
consistent with the presence of WR massive stars in similar 
young starburst knots (Lipari et al. 2003; 2004d).

Thus, for Mrk 231  we found a new example where WR stars are 
associated with a main complex of H {\sc ii} regions located
at the border of  a supergiant bubble (S1).
A similar result was found for all the main points of rupture of
the supergiant bubble of the nearby IR merger NGC 5514
(see Lipari et al. 2004d and the next sub-section).

\subsection{The OF in the strong complex of H {\sc ii} regions
in the external shell S1}
\label{results-6.5h2}

Hereafter,
we call {\it ``Aconcagua"} at this interesting complex of H {\sc ii}
regions, located at the south-west border of the supergiant shell S1
and associated with the strong knots K11, K12 and K14.
In our previous 3D kinematics study of Mrk 231 (based
on La Palma WHT+Integral spectroscopy for the central region
16.4$''\times$12.3$''$ $\sim$13 kpc$\times 10\,kpc$) we found
some evidence that this area could be associated to a main point
of rupture of the supergiant shell/bubble S1. In particular,
the H$\alpha$ isovelocity colour map for the main component of
the emission line (see Fig. 17, adapted from Lipari et al. 2005a:
their Fig. 7) shows two opposite lobes --blue and redshifted-- associated
with this area, at the border of the shell S1.

In this paper, we found for this complex of H {\sc ii} region Aconcagua:
(i) multiple OF components with high value of velocities (--400 km s$^{-1}$);
(ii) a  strong knot of [O {\sc iii}]5007 emission, which is
very close to K14;
(iii) the presence of high number of massive WR stars, in the knot K14--East.

It is important to remark, that previously
we also detected the presence of several knots with massive WR stars
in the supergiant bubble of NGC 5514 (and inside the areas of rupture
of the external shell).
For NGC 5514 we already proposed that a population of very massive stars
(like WRs) could be the source of multiple SN and HyN, and thus also the
source of the rupture of the bubble.

In Sections 12 and 10 we will present more detailed evidence that Aconcagua
 (K14+K12+K11) is probably associated with the areas of
ejection of the ionized gas and a main point of rupture of the more
external supergiant bubble S1.

The properties of the 4 bubbles and their origin will be
discussed in Section 13.3

\clearpage

\section{Mapping with GMOS-IFU the 
emission line ratios and the ionization structure}
\label{results-emlr}

Using the high spatial and spectral resolution  GMOS-IFU data (which
cover the main structures of the nuclear region and the supergiant
bubbles) we have studied in detail the ionization
and the physical conditions in: (i) the 4 supergiant shells; and
(ii)  the  nuclear narrow line region (NLR). This last point is
specially important since it is well known the absence of the
standard NLR in Fe II + IR + BAL QSOs
(see for details Lipari \& Terlevich 2006; Veron et al. 2006;
 Turnsheck et al. 1997; Lipari et al. 1994).

\subsection{Mapping with GMOS the nuclear ionization structure}
\label{results-emlr1}

Figs. 18 (a) and (b) show the 3D maps
(of $\sim$3\farcs5$\times$9\farcs0, $\sim$3 kpc $\times$ 7.3 kpc,
 with a spatial sampling of 0\farcs1)
of the emission line ratios
[N {\sc ii}]$\lambda$6583/H$\alpha$, and
[S {\sc ii}]${\lambda 6717 + 31}$/H$\alpha$.
These maps were constructed using the techniques described in Section 3
and are based on the main component of the emission lines.

Figs. 18a, b  show --with high spatial resolution-- very interesting features. We note the following:

\begin{enumerate}

\item
Coincident with almost the border of the more extended superbubble or
shell (S1),  both maps show arcs and knots with high values ($>$ 0.8) in the
[S {\sc ii}]${\lambda 6717 + 31}$/H$\alpha$ and [N {\sc ii}]/H$\alpha$
emission line ratios.
These arcs could be associated   mainly with shock processes at the
border of the supergiant  bubble S1.

\item
To the north-west, the GMOS mosaic
[N {\sc ii}]/H$\alpha$ map
shows  2 partial arcs (with several knots), which are probably associated
with shocks in the north-west part of the S2 and S3 supergiant bubbles.

\item
To the east of the GMOS mosaic, the
[S {\sc ii}]\,${\lambda 6717 + 31}$/H$\alpha$ map
depicts a clear arc (at r $= \sim$1$''$), which is probably associated
with shocks in the S3 superbubble.

\item
For the circumnuclear area where we measured the narrow emission line
components: specially in the region of the supergiant shells,
the [S {\sc ii}]\,${\lambda 6717 + 31}$/H$\alpha$ and
[N\,{\sc ii}]/H$\alpha$ emission line ratios
show high values ($>$ 0.5).

\end{enumerate}

\subsection{GMOS emission line ratios diagrams in the supergiant bubbles
and the NLR }
\label{results-emlr2}

Using the individual spectrum of this GMOS data,
an interesting  study was performed of the ionizing source at:
(i) the expanding supergiant bubbles, and
(ii) the nuclear and circumnuclear NLR.

Mainly, the log [S {\sc ii}]/H$\alpha$ vs. log [O {\sc i}]/H$\alpha$
emission line diagram was used. Since, this diagram
is an important tools for the study of OF process and
shocks, in dense medium (see Heckman et al. 1990; Canto 1984;
Binette et al. 1985; Shull \& McKee 1979; Dopita 1995;
Lipari et al. 2004a,b, 2005a).

Again, it is important to note that only in the very deep GMOS 
3D data (with 1 hs. of exposure time, see Table 1) the
spectra show very high quality: S/N $>$ 5, in the weak
narrow emission lines of the host galaxy. Which is
required in order to study the NLR and the OF process in the nuclear and
circumnuclear regions (using individual GMOS spectra;
which were obtained with lens of 0.2$''$).

Figure 19a shows this
log-[S {\sc ii}]/H$\alpha$ vs. log-[O {\sc i}]/H$\alpha$ diagram
for the 4 more external supergiant bubbles.
In Fig. 19a, the values of emission lines ratios (of the bubbles)
were obtained from Tables 8, 9, 10 and 11.
It is interesting to remark the following main points,

\begin{enumerate}

\item
Almost all the knots of the 4 external supergiant bubble
are located in the area of SNR + HH (i.e., the shocks
area), or in the transition/composite region between
SNR+HH and H {\sc ii} regions.
Thus in these areas the OF process play a main role.

\item
Only the knot S3a (K3 in Table 2 and Surace et al. 1998)
depicts emission line ratios typical of H {\sc ii} region.

\item
In the more internal supergiant bubble S4, the ELR of all
the knots  show a position close to the upper--right part
of this diagram (specially the knots located at the
south-west area). This fact is consistent with the presence
of strong [S {\sc ii}] and [O {\sc i}] emission; and thus
it is also consistent with shocks process of low velocities
($\sim$ 200 km s$^{-1}$; Heckman et al. 1990; Dopita
\& Southerland 1995; Binette et al. 1985; Shull 1980;
Shull \& Mc Kee 1979).

\end{enumerate}

It is important to study in detail also the narrow line emission in the
very nucleus, and the nuclear/circumnuclear regions. 
Tables 12, 13, 14, 15, 16 and 17 show the fluxes and ELR
for the very nucleus and for the nuclear area (for r $<$ 2.0$''$)
at the position angle PA $=$ 00$^{\circ}$. At this PA the OF process
is very strong/important, at low and large galactic scale
(Lipari et al. 2005a).
Figure 19b shows  
this log-[S {\sc ii}]/H$\alpha$ vs. log-[O {\sc i}]/H$\alpha$ diagram
for the very nucleus and the nuclear region at PA $=$ 00$^{\circ}$
(the values of these emission lines ratios were obtained from
Tables 16 and 17).
From this plot and Tables, it is interesting to remark the following points,

\begin{enumerate}

\item
In the  area close to the very nucleus (for r $<$ 0.7$'' \sim$ 0.6 kpc),
we detected similar properties to those observed in the more internal
superbubble S4: i.e., two similar narrow emission
line systems, with strong [S {\sc ii}] and [O {\sc i}] emission
and $\Delta$V $\sim$ --200 km s$^{-1}$.
Furthermore, even at the very nucleus these two emission lines systems were
detected.

\item
In the log [S {\sc ii}]/H$\alpha$ vs. log [O {\sc i}]/H$\alpha$ diagram
the ELR of the area close to the very nucleus (r $<$ 0.7$''$) also shows a
position close to the upper-right part of this diagram.
Again, this fact is consistent with the presence
of strong [S {\sc ii}] and [O {\sc i}] emission; and thus
it is consistent with shocks process of low velocities
($\sim$ 200 km s$^{-1}$; Heckman et al. 1990; Dopita \& Southerland 1995).
These results suggest that --in Mrk 231-- the NLR could be associated
mainly with the OF process.

\item
The ELR of the nuclear region (at PA $=$ 00$^{\circ}$, and for r $<$ 2.2$''$
$\sim$ 1.8 kpc)
are located in the log [S {\sc ii}]/H$\alpha$ vs. log [O {\sc i}]/H$\alpha$
diagram in the area of shocks: SNR + HH. However, in  the south nuclear
region (where we detected the blow-out phase of the bubble) the
ELR are located --again-- in the upper-right border of this diagram.
Which is consistent with the presence
of strong [S {\sc ii}] and [O {\sc i}] emission and shock process.

\end{enumerate}

\clearpage

\section{Mapping with GMOS-IFU the nuclear/circumnuclear A-type
stellar population}
\label{results-postsb11}

Adams (1972); Adams \& Weedman (1972); Boksenberg et al. (1977), and
others, detected the presence of an interesting stellar absorption line
system, in the nuclear and circumnuclear spectra of Mrk 231.
This stellar systems is observed with the GMOS data as strong absorptions
in the blue Balmer H lines, mainly in: H$\delta$, H$\gamma$ and H$\beta$.
From the study of the GMOS spectra of this A-type stellar population, new
interesting results were found:

\begin{enumerate}

\item
For H$\beta$  a range of equivalent width (EqW) of 7 -- 15 \AA, and
FWHM of 400 -- 470 km s$^{-1}$ were measured.
These absorptions were detected at almost all the field
covered by the GMOS mosaic (the nuclear area and the more extended
bubble S1).

These values (of H$\beta$) were obtained using  the wavelength
windows suggested by Gonzalez Delgado, Leitherer \& Heckman (1999).
Which will allow us to compare our results with their
 synthetic spectra of H Balmer and He I absorption lines (for
 starbursts and post-starburst galaxies).

\item
Using the lines H$\beta$ and H$\gamma$ a redshift Z $=$ 0.04255
(cz $=$ 12765 km s$^{-1}$) was measured, for this stellar
absorption system. Which is very close to that
obtained for the main component of the emission line.

\item
In addition, an interesting result was found when
the position of the strong H$\beta$ absorptions were plotted:
they are located close to the external border of the two
more external supergiant shells S1 and S2 (and thus with also
``arc--shape")

\end{enumerate}

Finally, a detailed study of the nuclear kinematics of this
A-type stellar population (using the  H$\gamma$ absorption line)
was performed. In the next Section
the main results of this study will be presented. In addition,
in Section 13.1, the properties of this stellar population will be
discussed. \\

\clearpage

\section{Mapping with GMOS-IFU the nuclear/circumnuclear kinematics}
\label{results-kinema13}

In this work we preset -for the first time- a study of the kinematics
of the stellar population of the nuclear region of Mrk 231 (plus the
kinematics of the high and low exitation ionized gas).
This study  was performed in two main steps:
first, the general properties of the  absorption lines of stellar
population and the emission lines were analysed. Using mainly
the corresponding velocity field (VF) maps and individual GMOS spectra.
Then for the observed stellar VF a  rotation curves (RCs) and the
nuclear mass were derived, for a radius r $<$ 0.7 kpc.

We already presented in previous Sections   the H$\alpha$ VF
for the central region of Mrk 231: 16.4$''\times$12.3$''$,  with a
spatial sampling of 0\farcs9 (Fig. 17; where also the area of the GMOS
mosaic is depicted).
This VF was obtained by Lipari et al. (2005a), from La Palma WHT +
Integral.
In the present paper this La Palma H$\alpha$--VF of Mrk 231 will be
compared with the GMOS kinematics data. Furthermore, this VF/figure
is important for the study of the rupture of the shell S1 (at the
HII region complex: Aconcagua) and for the extended OF process.

\subsection{Mapping with GMOS-IFU the ionized gas kinematics
([O {\sc iii}]$\lambda$5007 and H$\alpha$)}
\label{results-kig}

Fig. 20(a) and (b) show  --for the ionized gas--
the H$\alpha$ and [O{\sc iii}]5007 velocity field
maps for the nuclear and circumnuclear region of
$\sim$3\farcs5$\times$9\farcs0 ($\sim$3 kpc $\times$ 7.3 kpc),
 with high spatial resolution or sampling of 0\farcs1.
This map was constructed using the techniques described in Section 3,
and for the main component of the emission lines.
The errors vary from approximately
$\pm5\,$km~s$^{-1}$ in the nuclear and central regions (where the emission
lines are strong), to $\sim$10 km~s$^{-1}$ for the weakest lines away
from the nuclear areas.

The H$\alpha$ and [O{\sc iii}]5007 isovelocity colour maps (Figs. 20a, b)
show some interesting new structures:

\begin{enumerate}

\item
The H$\alpha$ VF in general shows redshifted velocity values in the
east area of the merger, and blueshifted values to the west. However,
very high blueshifted values were found in the south and south-west
regions.
Which is good agreement with the fact that these 2 regions (S and SW)
are both the area associated with  the rupture of supergiant
bubble S1 (i.e., associated with strong OF process).

\item
The [O{\sc iii}]5007 VF shows --in general- similar shape the  H$\alpha$
VF, but in the region of the nucleus
($\sim\Delta\alpha =$ 0, and $\sim\Delta\delta =$ +2$''$)
this [O {\sc iii}] VF depicts very interesting sub-structure.
In particular, we found filaments structures in the north and north-east
direction, which are in good agreement with the near IR 3D spectroscopic
study reported by Krabbe et al. (1997).

\item
The highest blueshifted values were found at the south-west border of
both GMOS VFs, and close to the position of the supergiant bubble S1.
Probably, these kinematics structures --or lobes-- are associated with knots
 of rupture/blow-out process (i.e., K11+K12+14).

\end{enumerate}

Therefore, the  GMOS-IFU and Integral 3D spectroscopic studies 
suggest that the complex of the knots K11 + K12 + K14 is a main
point of rupture --or blow-out-- of the supergiant shell S1.

\subsection{Mapping with GMOS-IFU the Inter Stellar Medium kinematics
(using the galactic Na ID absorption line)}
\label{results-kism}

Fig. 20c show the GMOS VF map obtained from the galactic Na ID
absorption line: i.e., the Na ID detected at the systemic
velocity of the merger (cz $=$ 12654 km s$^{-1}$).

From this figure 20c it interesting to note the following points.

\begin{enumerate}

\item
In all the nuclear and in the nearest circumnuclear area
--of the very nucleus--
there is a clear OF process: with two strong blueshifted lobes.
This result is clearly consistent with the
previous studies of the OF process (in the nuclear region of Mrk 231),
which detected extreme velocity OF of ionized gas.

\item
There is also a weak blueshifted lobe (and filaments) in the north
and north--west region, at a radius r $\sim$ 2$''$.
For this area Krabbe et al (1997) reported
several OF and filaments.

\end{enumerate}

\subsection{Mapping with GMOS-IFU the stellar kinematic and the nuclear
rotation curve}
\label{results-kstellar}

Fig. 20d shows the map of the stellar velocity field
from the H$\gamma$ absorption line, and
for the nuclear region of
$\sim$3\farcs5$\times$5\farcs0 ($\sim$3 kpc $\times$ 4 kpc),
 with a spatial resolution/sampling of 0\farcs1.
We note that the field of this figure is only for the ``nuclear"
GMOS-IFU frame, and it do not cover  all the GMOS mosaic,
since the H$\gamma$ absorption is very week in almost all the
GMOS frame of the southern arc.

In this nuclear stellar VF is evident the presence, close to the
very nucleus (at 0.8--0.9$''$ and at PA = 00$^{\circ}$), of two
symmetric lobes: i.e.,
a blueshifted circular lobe located to the south (of the very nucleus) and
a redshifted circular lobe to the north. This is the typical structure of
circular motion around the very nucleus. 

Thus, from this stellar VF and for the region close to the very nucleus
(r $<$ 1.4 kpc) a  rotation curve was constructed, using the software
ADHOC and the technique already described by Lipari et al. (2004a).
Specifically, the RC was obtained
from the extraction in the VF of an angular sector of $\pm$20$^{\circ}$
around the line/axis  north-south (at PA 00$^{\circ}$).
The shape of the observed ``nuclear" RC of Mrk 231
--with a monotonic increase in the more external nuclear regions,
and after reaching a maximum, a symmetrical fall outwards--has been
called a ``sinusoidal'' shape (for details, see Rubin \& Ford 1983;
Zepf 1993; Mendes de Oliveira et al. 1998; Lipari et al. 2004a, 2000).
In the advanced mergers NGC\,3256 and NGC 2623, we already found similar
sinusoidal RCs, but at larger scale (i.e., for all the radius of the
galaxy/merger).
We suggested that this sinusoidal feature could be a common characteristic of
advanced major merger systems, and probably associated with OF
in the more external regions (L\'{\i}pari et al. 2000a, 2004a,c).

From the RC (Fig. 20e)  we derive a Keplerian mass, inside
a radius $r = 700$ pc,
M$_{dynam}$(700 pc) $\sim$ (0.5 $\pm$0.2) $\times$ 10$^{10}\ M_{\odot}$.
In order to confirm this result and to analyse more physically the RC,
we fitted this nuclear RC with several well known mass distribution models:
following the step and technique described in detail in our study of the
RCs of NGC 2623 and NGC 3256 (see Lipari et al. 2004a,c, 2000).
The best fit to the nuclear rotation curve  was obtained with  a
Plummer-Kuzmin law (Binney \& Tremaine 1987), which correspond to
a spherical or disk mass distribution.
For the Plummer-Kuzmin potential, we have used the relation,

\begin{equation}
 \Phi(r) = {- G M_T \over \sqrt{r^2 + r_0^2}},   
\end{equation}

where $G$ is the gravitational constant,  $M_T$ is the total mass and $r_0$
is the scale length corresponding to
$r_{\rm max}$/$\sqrt{2}$,  $r_{\rm max}$ being the turnover radius of the RC.
The circular velocity in the equatorial plane is, in cylindrical coordinates,

\begin{equation}
 V_{\Theta}^2(R,0) =  { G M_T R^2 \over [R^2 + R_0^2]^{3/2}},  
\end{equation}

The best fit was obtained with
M$_T = M_{TOT. NUC.} = 0.9 \times 10^{10}\ M_{\odot}$; and with the following
geometry PA$_0 = 00^{\circ}\pm5^{\circ}$, $i = 45^{\circ}\pm5^{\circ}$
(a mean value), and $V_{\rm sys} = 12650 \pm$ 10 km\,s$^{-1}$.
The value of mass (derived using a Plummer-Kuzmin law) is consistent
with  the value obtained for a Keplerian mass.

Furthermore, it is important to remark
that our derived values of the nuclear mass (obtained in this paper and for
the inner 0.7 kpc) are within the errors similar --or slightly lower-- that
the value obtained by Downes \& Solomon (1998) for the dynamical mass,
inside of the radius $r = 0.5$ kpc: 1.2 $\times$ 10$^{10}\ M_{\odot}$.
Another interesting point --about this last work-- is the fact that the
radius of $\sim$0.5 kpc (used by Downes \& Solomon 1998, for their study
of the mass in the nuclear region of Mrk 231; using CO observations and
models), is the radius of the starburst disk detected using
CO interferometric-IRAM data.
Thus, there is a good agreement between the GMOS and IRAM high spatial
resolution studies, performed at different wavelength regions and
using different techniques.
More specifically, in the present work a good agreement was found between
the kinematics results obtained: at optical and millimetre wavelengths and
using high resolution GMOS 3D spectroscopy and IRAM
interferometric techniques. \\

\clearpage

\section{DISCUSSION OF THE GMOS-IFU RESULTS}
\label{discussion}

\subsection{Decoupling with GMOS-IFU the very young nuclear
starburst from the QSO}
\label{compositenuc}

\subsubsection{The GMOS nuclear extreme blue/starburst component
(and the composite evolutionary model for BAL QSOs)}
\label{bluecompocomposite}

In this paper, using the GMOS-IFU spectra and GALFIT 3D Model we found:
(i) in the pure host galaxy spectrum an extreme nuclear starburst
component (for the first time, at optical wavelength), which was detected
mainly as a very strong increase in the flux, at the blue region;
(ii) 3D GMOS individual spectra and maps  confirm the presence of an extreme
blue/starburst component, which was detected in a nuclear dusty ring or  
toroid.

More specifically,
in Section 6 we found in the map of the nuclear blue continuum (or colour
index, Fig. 8b) two clear, symmetric and strong peaks, which are located
at 0\farcs3 to the south and north, of the very nucleus.
This position of the peak is coincident with
the location of the ring S5.
Thus, these two peaks are probably positioned inside of the dusty shell/ring S5.
In addition, different previous studies, performed at radio wavelength,
already proposed that in this area there is a disk of molecular gas, with
extreme star formation (SFR) rate of $\sim$100--200 M$_{\odot}$ yr$^{-1}$
(Bryan \& Scoville 1996; Downes \& Solomon 1998; Carrilli et al. 1998;
Taylor et al. 1999).

On the other hand, in this region (of the blue and red peaks) the individual
GMOS spectra show strong and narrow multiple emission line components
(with OF), specially in the lines [O {\sc ii}]$\lambda$3727; plus stellar
narrow emission and absorption in the IR Ca {\sc ii}
triplet. These are clear signature of very young stellar population.
The presence of two strong narrow [O {\sc ii}]$\lambda$3727 components
could be associated mainly with young H {\sc II} regions with strong OF.

The strong narrow emission plus absorption in the IR Ca
II$\lambda$8500 triplet (detected in the symmetric areas of the red and blue
continuum peaks) is clearly associated with the peak of
RSGs activity, with age of: 8 Myr $<$ age $<$ 15 Myr (in metal-rich stellar
populations; see Terlevich et al. 1991; Lipari \& Terlevich 2006).
The red continuum peak could be associated with a very dusty region
of the same starburst of the blue continuum component.
Furthermore, probably both areas (of the blue and red continuum peaks)
are  associated with the shell/ring S5.

It is important to remark, 
the south nuclear and circumnuclear  area (with blue continuum) is coincident
with the region where we previously suggested that the nuclear galactic wind
in Mrk 231 is cleaning the nuclear dust.
Specifically,
we suggested that the more external supergiant bubble S1 is
in the blow-out phase: i.e. cleaning all the south area of the whole merger.
Thus this fact is probably allowing to see the extreme nuclear
starburst.

It is interesting to remark that the derived dynamical time for
the more external supergiant bubble S1: $\sim$ 5 $\times$ 10$^{6}$ yr,
suggest that this bubble could be generate in the first phases
this very young starburst, which is probably associated with the interaction
between the star formation process  with the AGN + accretion disk.
In addition, it is interesting to note that the
last/young starburst activity detected in this
merger must have occurred very late in the history of the interaction,
since the age of the starburst is only 5--20 $\times$ 10$^6$ yr.
This is in agreement with the results published by Mihos \& Hernquinst
(1996) for models of starburst in disk/bulge/halo major mergers process
(their Figs. 4 and 2), where extended massive star formation processes
occurred mainly late in the history of the interaction.

Therefore,
these new GMOS results obtained in this paper: the detection of a very young
nuclear starburst in Mkr 231 (mainly by the presence of a strong blue
continuum flux in the nucleus of the host galaxy and the detection
of the near-IR Ca {\sc ii} triplet in
emission plus absorption, probably in a toroid close to the very nucleus)
clearly confirm --and are in good agreement with-- the proposed
{\it evolutionary and composite model} for  composite
BAL + Fe {\sc ii} + IR QSOs.

\subsubsection{The GMOS data of the extended  post-starburst
(or A-type stellar population) }
\label{post-sb}

In Section 11, for the absorption of H$\beta$  a range of equivalent
width  of 7 -- 15 \AA, was obtained.
This result was analysed using the evolutionary stellar population
synthesis models of Gonzalez Delgado, Leitherer \& Heckman (1999).
Specifically, we compared the H$\beta$ EqW observed for Mrk 231 with
the grid of EqW of the models (for H Balmer and He I lines).
The used  model correspond to a cluster with: instantaneous burst,
solar metallicity and  Salpeter IMF, between
M$_{low} =$ 1 M$_{\odot}$ and M$_{up} =$ 80 M$_{\odot}$.
From this study, a range of age of 30 -- 500 Myr was obtained
for the A-type stellar population of Mrk 231 and a T$_{eff}$ of
9000 K.
This result is in good agreement with the idea that this A-type stellar
population could be associated with the first phases of the merger process
of Mrk 231. Since, for example the time scale of mergers with prograde
orbit is: 0.5 $\times$ 10$^{9}$ yr (Barnes 1992; Noguchi 1991).

An interesting result obtained in Section 11 is the fact
 the regions with strong H Balmer absorption lines show 
``arc shape" morphology (which are located very close to the external
border of the supergiant shells S1 and S2). A simple explanation for this
result could be that the OF process --in these shells-- is cleaning the
nuclear dust.
Thus, this cleaning process is probably allowing to see more clearly
(and strong) the absorptions of this A-type stellar population,
close to the external parts of the expanding shells.


\subsection{Decoupling with GMOS-IFU the BAL systems: the jet-wind
wind nature of the BAL {\sc i} and {\sc ii} systems (plus the explosive
nature of the BAL {\sc iii})}
\label{extendedbal}

\subsubsection{The AGN jet-wind and the extended nature of the BAL Systems
{\sc i} and {\sc ii}}
\label{extendedbal2}

\noindent
{\bf Main predictions of the AGN and composite Models for BALs}

In general, there are two groups of models about the structure and
dynamics of the gas near the core  of AGNs. Specifically in the
broad absorption line region (BALR) and the broad emission line region (BELR).
In these models the gas may exist as:
(i) continuous winds: spectral analysis of Arav et al. (1994, 1997, 1998,
2001, 2005)
seem to show that continuous winds might be better suited to explain
high resolution spectra spectra of BALR and BELR.
(ii) discrete clouds: the idea that gas is partitioned into discrete
clouds is the more traditional approach to BELR and BALR (see Everett, Konigl,
Kartje 2000; Bottorff et al. 1997).

The two main AGN theoretical models for BALs (Jets and Accretion Disks)
predict some clear effects that could be detected with high spatial and
spectral resolution observations (and with high quality/SN spectra).
The deep GMOS-IFU data presented in this paper are probably one of the
best and complete set of data available to date for this purpose.
Furthermore, these GMOS data were obtained for the nearest BAL QSO.

Specifically, these two model predict some very simple and clear effects
(see for a discussion of this point Punsly \& Lipari 2005;
Proga et al. 2000; Punsly 1999a,b; Murray et al. 1995).
Here, we present a summary  of the more clear prediction of these two
AGN models:

\begin{enumerate}

\item
{\it Jet Model for BAL:}
This model predict that the jet wind is collimate,  bipolar and
aligned with the Jet direction. Thus any possible elongation will
be detected in the jet direction.

It is important to note, that
in order to detect any elongation in a BAL system it is required
an extended nature of the studied BAL, in a scale similar to the
spatial resolution of the instrument (for GMOS $\sim$ 0.4--0.5$''$).
Different effects could explain the extended nature of some BALs
(Punsly \& Lipari 2007, in preparation).

\item
{\it Accretion Disk Model for BAL}
In this second model the accretion disk wind is located close to the
plane of the disk, i.e. for AGNs with detected Jets, the wind associated
with accretion disk is located perpendicular to the direction of the Jet.
Thus any possible elongation in the BAL will
be detected perpendicular to the jet direction.

\end{enumerate}

It is interesting to note that in the composite + supergiant shell scenario
for BALs, some elongations could be also generated. Specially, the
elongations could be observed in the case of collimate/bipolar expansion of
the galactic wind + supergiant shells.
This type of OF could generate elongations at the position angle
of the direction/axis of the bipolar galactic wind (or hyperwind) +
supershells.\\


\noindent
{\bf The jet-wind nature of the BALs {\sc i} and {\sc ii}
systems}

In the present work,
we have studied (using high resolution very deep 3D GMOS spectra)
all the optical  ``absorption" lines associated with the BAL
systems {\sc i} and {\sc ii}.
The deep 3D GMOS spectra and maps clearly show that
the BAL systems {\sc i} (in the Ca {\sc ii} K$\lambda$3933 absorption map)
and the BAL system {\sc ii} (in the Na ID$\lambda$5889-95 absorption map)
are clearly elongated at the position angle close to the radio jet PA.
Which strongly suggest that the BAL systems {\sc i} and {\sc ii} are
``both" associated
with the radio jet, and supporting the bipolar jet-wind model for some BALs.
Thus, this new study (based in deep GMOS-IFU spectroscopy)
is in excellent agreement with the previous 3D Integral spectroscopy study
of the H$\alpha$ emission bump.

In addition,
the very deep 3D GMOS spectra (Fig. 10) and maps (Fig. 11) clearly show
the extended nature of the BAL system I:
reaching $\sim$1.4--1.6$''$ $\sim$1.2--1.3 kpc, from the nucleus.
Which are also in excellent agreement with the extended nature found
previously (from the 3D Integral study of H$\alpha$ bump)
of the BAL-I. Furthermore,
de Kool et al. (2001, 2002) found similar results --extended
nature of two BAL systems-- in their Keck high resolution spectroscopic study
of BALs in the QSOs FIRST J104459.6+365605 and FBQS 0840+3633.
They found that the distances between the AGN and the region where the
OF gas generate the BAL line are $\sim$700 and $\sim$230 pc,
respectively.
Therefore, for Mrk 231 (BAL system--I), FIRST J104459.6+365605 and
FBQS 0840+3633 the distances found between BAL  forming regions and the
continuum source (AGN) are large,  $\sim$200--1400 pc. BALs
are generally thought to be formed in OF at a much smaller distance
from the nucleus (see for references de Kool et al. 2001).

\subsubsection{The explosive nature of the BAL {\sc iii} System }
\label{extendedbal4}

A very important point about the explosive scenario for Mrk 231 is the
detection (together with the multiple expanding shells) of an exponential
fall in the variability of the short lived  BAL {\sc iii} Na ID system.
In this paper, this exponential fall in the  BAL {\sc iii}
Na ID was confirmed (for almost all the period in which this system
appear, 1984--2005), using very deep 2005 GMOS IFU spectra.
The origin of this exponential fall in the BAL {\sc iii} system could be
explained, mainly in the frame work of an {\it extreme explosive event},
probably associated with hypernova explosions.
An explosive scenario for the origin of the BAL {\sc iii} system could
explain also the presence (in the nucleus) of multiple concentric
expanding superbubbles, with circular shape.

\subsubsection{Mrk 231 and the rare class of Fe II and Low ionization
BAL-QSOs}
\label{extendedbal2}

A very interesting  point is the fact that
Mrk 231, FIRST J104459.6+365605 and FBQS 0840+3633
(which shows extended BAL systems)
are all member of the rare class of low ionization BAL QSOs.
Furthermore, these 3 QSOs are also member of the  ``very" rare sub-class of
Fe II low ionization BAL QSOs with very strong reddening in the
UV continuum. In particular, for Mrk 231 Lipari et al. (2005a, their
Fig. 11a) clearly shows -in the UV spectrum- the presence of strong
absorption in the Fe II and Mg II lines; which are the standard lines
that define the  Fe II low ionization BAL QSO sub-class.

We already suggested that low --and specially Fe II low-- ionization BAL QSOs
are young and composite/transition QSOs: in the phase that the OF process
with galactic winds, expanding supergiant bubbles, and exploding HyN are
clenenning the nucleur dust (which is generated, at least in part by the
extreme nuclear starburst). Thus, the strong reddening in the
UV continuum is also associated with their composite nature.
In addition, Lipari (1994) and Lipari et al. (2005a) found -in the IR
colour-colour evolutionary diagram- a clear sequence of
transition BAL + Fe II + IR QSOs. Which are -almost all- member of the
rare class of {\bf low ionization BALs QSO} (like Mrk 231; IRAS 075988+6508;
IRAS/PG 17002+5153, etc).

Very recently,  from a study of a very large sample of 37644 Sloan Digital
Sky Survey (SDSS)  QSOs, from the 3rd. Data Release (DR3) and
for all redshift (in the range: 0 $<$ z $<$ 5): White et al. (2006) found
that the radio properties of the rare class of low ionization BALs QSOs are
different to the group of non-BAL QSOs + high ionization BAL QSOs, at all
redshift. This result could
be explained mainly in the frame work of an evolutionary unified scenario
for BAL QSOs, in close agreement with the model proposed by
Lipari \& Terlevich (2006): the low ionization BAL QSOs are young, transition,
and composite QSOs in the phase of a strong/composite OF, with supergiant
expanding shells (probably associated with exploding HyN and Jets); which
is cleaning the nuclear dust.


\subsection{Decoupling with GMOS-IFU the nuclear outflow and the expanding
superbubbles,  in Mrk 231 }
\label{msbubble}

In the present paper,
we have confirmed with high spatial and spectral resolution GMOS data
the presence of multiple concentric expanding supergiant bubbles/shells,
with: centre in the nucleus and highly symmetric circular shape.
These shells could
be associated mainly with giant symmetric explosive events.
These explosive events could be explained in a composite
scenario, where the interaction between the starburst and the AGN
could generate giant explosive events: i.e., HyN
(Collin \& Zahn 1999; Artymowicz et al. 1993).
Furthermore, 
we derived for the kinetic energy of the OF of the shell
S1, E$_{\rm KIN OF-S1} \sim$  2.0 $\times$ 10$^{54}$ erg
(Lipari et al. 2005a). This very high level of kinetic energy,
obviously required the presence of multiple SN events,
or an unusual type of ``giant SN" or hypernovae
(with E$_{\rm KIN HyN} >$  10$^{52}$ erg; see Nomoto et al. 2006,
2007, 2004). Which is also in good
agreement with the extreme SB and composite scenario for the nucleus
of Mrk 231.

In this work we have also presented GMOS morphological and kinematics
evidence that that the main complex of H {\sc ii} regions (K11+K12+K14:
``Aconcagua") could be associated to a  main point of rupture of the
border of the bubble S1 (with a large number of massive/WR stars).
In the next sub-Sections this point is discussed in detail.

\subsubsection{GMOS evidence of the rupture of the external supergiant shell}
\label{dis-gwrupture}

In Sections 9, 10 and 12 we have studied the 3D GMOS spectra and
images of the main knots of the multiple expanding nuclear bubbles.
In particular,
we found strong multiple [O {\sc iii}]$\lambda$5007 emission and
Wolf Rayet features in the main complex (K11+K12+K14: {\it ``Aconcagua"})
of the more external supergiant bubble S1.
These results are in good agreement with those obtained previously,
in the sense that: the H {\sc ii} region complex ``Aconcagua"
is probably associated with a main point of rupture of the bubble.

More specifically,
in this work we found the signature of a large number of WR and
very-massive stars in the strong  knot K14-East
associated with a main complex of H {\sc ii} regions, located at
the south--west border of the more external supergiant bubble S1.
It is important to remark, that
populations of massive stars are the main progenitors of ``multiple"
core--collapse SNe/HyN. Which
 are the main objects capable to generate the rupture phase of the
 galactic bubbles (Heiles 1979; Norman \& Ikeuchi 1989).

Recently, a similar results was found for all the main points of
rupture of the supergiant extranuclear bubble in NGC 5514: i.e.,
the presence of WR stars in all the knots of
rupture  of the  supergiant shell. Furthermore,
for NGC 5514 we already proposed that a population of very massive stars
(like WRs) could be the source of multiple SN and HyN, and thus also the
source of the rupture of the bubble.

Therefore,
using deep GMOS-IFU high spatial resolution 3D spectra, we found
--in this paper and for the distant IR merger Mrk 231--
very similar results to those previously detected in the nearby merger
NGC 5514 (this last object is considered the proto-type of a IR merger
with a supergiant bubble in the rupture phase).

\subsubsection{GMOS data and the nature of the nuclear multiple
expanding suprergiant shells}
\label{dis-gwm231}

In Sections 9, 10 and 12 we found new interesting results about the
physical and kinematics properties of the nuclear expanding bubbles.
Here, these properties and the nature of these
five nuclear expanding supergiant bubbles/shells are discussed.

Specifically, we remark the following main points:

\begin{enumerate}

\item
{\it Supergiant bubble/shell S1:}

In this paper we found strong evidence that this supergiant
bubble is in the blowout phase. Furthermore, the typical
kinematics, morphological and physical properties of a point
of rupture of the bubble were found (mainly in the knot K14-East,
of this relatively distant IR merger).

This region of rupture of the bubble S1 (south west area) and also several
knots in the 4 external supergiant bubbles show clear blueshifted
emission, in the OF and --even-- in the main components.
With a value of OF, of
$\Delta$V $\sim$ --400 km s$^{-1}$. \\

\item
{ \it Supergiant bubble/shell S2:}

The main/strong knots (of this shell) are all located in the south west
region with typical emission line ratios of LINER,
associated with shocks of low velocity. \\

\item
{\it Supergiant bubble/shell S3:}

The  knots of this shell show composite values of ELRs, between
LINER and H {\sc ii} regions. In addition, the knot S3a (also called K3)
shows typical values of emission line ratios of
H {\sc ii} regions.

This last result is
in good agreement with that obtained for the near area E1 (where we
detected  very strong H$\alpha$ emission),
which also shows typical  emission line  ratios of H {\sc ii} regions. \\

\item
{\it Supergiant bubble/shell S4:}

In this internal bubble --and close to the very nucleus,
for r $<$ 0.7$" \sim$ 0.6 kpc-- two similar narrow emission
line systems were detected, with strong [S {\sc ii}] and
[O {\sc i}] emission (with $\Delta$V $\sim$ --200 km s$^{-1}$).
These results are consistent with ionization by OF + shocks
of low velocity and in a dense ISM. \\

\item
{\it Ring or toroid S5:}

In this paper we found evidence that this ring is probably
associated with a toroid of very young and strong star formation 
process of 8 $<$ age $<$ 15 Myr (see for detail Sections
6 and 7). \\

\end{enumerate}

\subsection{GMOS data and the  nature of the nuclear NLR and BLR
(in Mrk 231)}
\label{dis-nlr}

In general, the main results obtained in Sections 10 and 12 clearly show that
the kinematics, morphology and physical
properties of the multiple narrow emission lines systems (in the
very nucleus and the nuclear region, for r
$<$ 2.2$" \sim$1.8 kpc) are all consistent with an ionization process
generated by  the OF  with low velocity shocks. In addition,
only few nuclear area were detected, where the dominant ionization
could be associated with  H {\sc ii} regions.

More specifically, the following results are consistent
with a nuclear NLR associated mainly with the OF gas and
with an ionization process by shocks,

\begin{enumerate}

\item
In the  area close to the very nucleus (for r $<$ 0.7$" \sim$ 0.6 kpc),
two similar narrow emission
line systems were detected, with strong [S {\sc ii}] and [O {\sc i}] emission
and $\Delta$V $\sim$ --200 km s$^{-1}$.
In addition, even at the very nucleus these two emission lines systems were
observed.

\item
In the log-[S {\sc ii}]/H$\alpha$ vs. log-[O {\sc i}]/H$\alpha$ diagram
the ELR of the area close to the very nucleus (r $<$ 0.7$''$) also show a
position close to the upper-right part of this diagram
(specially in the south area).

This fact is consistent with the presence
of strong [S {\sc ii}] and [O {\sc i}] emission; and thus
it is consistent with shocks process of low velocities
($\sim$ 200 km s$^{-1}$; Heckman et al. 1990; Shull \& McKee 1979;
Dopita \& Southerland 1995).

\item
All the ELR for the nuclear region (at PA $=$ 00$^{\circ}$, and for
r $<$ 2.2$''$ $\sim$ 1.8 kpc)
are located in the log-[S {\sc ii}]/H$\alpha$ vs. log-[O {\sc i}]/H$\alpha$
diagram in the shocks area: of SNR + HH. In  the south nuclear
region (where we detected the blowout phase of the bubble) the
ELR are positioned --again-- in the upper-right border of this diagram.

\end{enumerate}

Therefore, these results suggest that --in the nuclear and circumnuclear
region Mrk 231-- the NLR could be associated mainly with the OF process.
It is important to remark, that even in dusty IR Galaxies we have found
several standard NLRs, in which the nuclear emission line ratios are all
consistent with ionization by the Seyfert/AGN activity.
For example in IRAS 15480--0344 (Lipari et al. 1991) we detected
a typical standard/Seyfert NLR. In addition, new detailed high resolution
narrow band images and spectra --of this extended NLR-- show only
ionization associated with the Seyfert activity (Tsvetanov 2007, private
communication). However, this is not the case for the NLR of
Mrk 231, where we did not detect any ionization in the NLR
associated with the Seyfert activity.
For Mkr 231, the emission line ratios of the NLR are clearly consistent with
a ionization process dominated by the OF + Shocks events.

On the other hand,
we already proposed that even the broad line  {\it emission} region of
Mrk 231 could be generated by an OF process. In particular,
we  showed that explosive and OF events (more constant that a
single and standard SN) could generate the unusual BLR spectrum  observed
in the very nucleus of Mrk 231 (see for details Lipari et al. 2005a). \\

\clearpage

\section{The composite hyper--wind model for Mrk 231,
 BAL + IR + Fe {\sc II} QSOs and Ly$\alpha$ blobs (works in progress)}
\label{evolutionqso3blob15}

In order to analyse the extended nature of the OF of Mrk 231 (and
BAL + IR + Fe II QSOs), it
is important to remark two previous studies at radio wavelengths,
which are important for the discussion of the extended and composite
OF process. In particular:

\begin{enumerate}

\item
VLA images at 4.9 Ghz (Baum et al. 1993) and at 1.5 Ghz (Ulvestad et al.
1999a) show a very large structure of $\sim$35--50 kpc,  elongated
in the direction north-south.

\item
VLBI images at 1.7 Ghz (Neff \& Ulvestad 1988; Lonsdale et al. 2003) and
VLBA images at 2.3 Ghz (Ulvestad et al. 1999a) show a triple
structure with a central unresolved core and two symmetric resolved lobes;
with a total extension of $\sim$40 pc.
This radio structure is elongated also in the direction north-south.

\end{enumerate}

These results clearly suggest that these extended structures
(from scale of few parsec to 50 Kpc) are probably associated
with the main components of a composite OF process:
the blowout phase of the more external bubble S1
(at PA = 00$^{\circ}$ and at large/kpc scale).
These results (for Mrk 231) are in good agreement with those
obtained previously for several BAL + IR + Fe {\sc ii} QSOs:
like  IRAS04505-2958 (see Lipari et al. 2007a,b).

\subsection{Composite hyper--wind model for BAL + IR + Fe {\sc ii} QSOs}
\label{hypwind2}

In several  BAL + IR + Fe {\sc ii} QSOs we detected extended OF processes
(similar to the extended Ly$\alpha$ blobs found at very high redshift;
Steidel et al. 2000)
which are probably associated with the composite nature of the very
nucleus of QSOs/AGNs (Lipari et al. 2500a, 2003). In particular,
Lipari et al. (2005a) proposed a {\it composite hyper--wind
scenario} in order to explain the very extended blob/shell --of $\sim$30
kpc-- found in the new BAL QSO IRAS04505-2958.

In addition,
in the study  of IRAS 07598+6508 and PG 1700+518,
Lipari (1994) already suggested that nearby and high redshift
{\bf low--ionization BAL QSOs} could be explained by a violent ejection
during the first onset of the QSO activity, similar to a  ``giant
SN explosion" (previously proposed by Hazard et al. 1984).
This approach is very similar to that suggested  for Mrk 231,
where ``multiple nuclear expanding super--bubbles"  were found,
associated --in part-- with giant SN or hypernova explosions.
Therefore, it is important to study the possible role of composite
hyper--wind/OF in the evolution of SMBHs and QSOs.

It is important to remark that very recently it was confirmed
our suggestion of the existence ef {\it extreme type IIn  SN explosive
events} associated with very masive progenitors, like Eta Carinae
(Lipari \& Terlevich 2006).
In particular, the discovery of the SN 2006gy (in NGC 1260,
Smith et al. 2006), that reached a peak of absolute magnitud of --22,
and remain brigther than --21 mag for about 100 days.
Thus, this type IIn SN is the most luminous SN ever recorded
powered by the death of an extremely massive star like Eta Carinae,
and shows a good agreement with one of the main suggestion of
the evolutionary model for composite AGNs (Lipari \& Terlevich 2006).
Furthermore,
Smith et al. (2006) even proposed that giant SN explosions from
very massive progenitors could be more numerous in Population III stars
(in young objects and in the early universe) than previously belived.
Finally, it is important to remark that this SN 2006gy show very strong
 Na ID and H \& K Ca II absorption lines (i.e., very similar to those
 of Mrk 231).

\subsection{Composite hyper--wind model for Ly$\alpha$ blobs}
\label{hypwind3}

The ionizing radiation from the newly formed young stars should lead to
prominent Ly$\alpha$ emission due to recombination of the hydrogen in
the ISM. Thus, extended Ly$\alpha$ emission could be an important
spectral signature of young and composite systems, specially at very
 high z (Terlevich et al. 2007, in preparation).

Recently, we have started a study of 3D spectroscopic data of high
redshift Sub-mm and Radio BAL-QSOs, using Gemini+GMOS and ESO VLT+VIMOS.
Specifically, we have already performed a detailed study of Gemini
GMOS-IFU data of the low z BAL QSO IRAS 04505-2958 and high redshift
Sub-mm SDSS BAL QSOs (Lipari et al. 2007b, in preparation).
The results of this work is the second paper
of our Gemini GMOS IFU programme of BAL QSOs. In general, we found that
IRAS 04505-2958 and Mrk 231 show  similar extended OF process (of
100 and 50 kpc, respectively). Even
both QSOs have   ``relatively narrow" --or nini/associated-- BALs
(Lipari et al. 2007a,b).

Therefore, we are studying if extreme OF associated with  giant
explosions and hypernovae (plus AGNs jets) could generate
 BAL systems and extended blobs (Lipari \&
Terlevich 2006; Lipari et al. 2007a, 2005a, 2003; Punsly \& Lipari 2005;
Reuland et al. 2003; Bond et al. 2001; Tenorio-Tagle et al. 1999; 
Guillemin \& Bergeron 1997; Dey et al. 1997; Lipari 1994).  
The main goal of this study is to test our proposed
 composite hyperwind scenario for some BAL QSOs at low and high redshifts
(Lipari et al. 2007a,b; Lipari \& Terlevich 2006; see also
Magain et al. 2005).

\clearpage

\section{The evolutionary end of Mrk 231, elliptical and QSOs
(and future works)}
\label{evolutionqso4end16}

The GMOS results obtained for Mrk 231 (combined with theoretical and
observational studies, for mergers with OF)
suggest that extreme starbursts and extreme galactic
winds play an important role in galaxy/QSO evolution (see Larson 2003,
1999, 1998; Bromm \& Loeb 2003; Bromm, Coppi, \& Larson 1999;
Scannapieco \& Broadhurst 2001; Lipari \& Terlevich 2006;
Lipari et al. 2007a).

On the other hand,
several results (for the merger Mrk 231) suggest that the nuclei of the
colliding galaxies have coalesced into a common nucleus,
and that the merger is in a very advanced phase: a relaxed
system probably evolving into an elliptical galaxy
(see Soifer et al.\ 2000; Quillen et al. 2001; Lipari et al.\ 1994;
Condon et al.\ 1991; Hamilton \& Keel 1987).
Thus, a very interesting point is to follow this evolutionary study of
Mrk 231 and similar evolving elliptical galaxies with composite and
extreme OF (observed mainly as extreme BAL + IR + Fe {\sc ii} QSOs),
 even to the end phase, of their evolution.

It is important to note,
that in the last years several possible \emph{links} between \emph{mergers,
starbursts, IR~QSOs and elliptical} have been proposed. Specifically,
L\'{\i}pari et al. suggested the following evolutionary--links:

Merger{\bf /s} $\to$  extreme starburst + galactic-wind
(inflow + outflow) $\to$  IR + Fe\,{\sc ii} + BAL composite/transition QSOs
$\to$ standard QSOs and elliptical $\to$ galaxy remnants.\\

In this evolutionary sequence a main and interesting step is end phase
of the evolution of the host galaxies + QSOs. 
We have started observational and theoretical studies in order
to analize if extreme and explosive OF process --in composite
BAL + IR + Fe II QSOs-- could be associated  with 3 main step in
the evolution of QSOs and their host galaxies.
In particular,
we are studing the role of explosive events in:

\begin{enumerate}

\item
To stop the accretion process in QSOs/SMBHs.

\item
The formation of satellite and companion galaxies (by
explosions).

\item
To define the final mass of the host galaxies, and even if the
explosive nuclear outflow is extremelly energetic, this process
could disrupt an important fraction (or even all) of the host galaxies.

\end{enumerate}

Therefore, giant QSOs explosions is an interesting process
in order to consider as the base for a first model of galaxy end.
Our observational GMOS-IFU results for Mrk 231, IRAS 04505--2958,
and others BAL + IR + Fe II QSOs, plus the theoretical works performed by
Ikeuchi (1981); Ostriker \& Cowie (1981); Berman \& Suchkov (1991) show
a good agreement with explosive models for the formation and end
of some type of galaxies (which are associated mainly with explosive
BAL + IR + Fe II QSOs; see for details Lipari et al. 2007a,b).

\clearpage


\section{Summary and Conclusions}

In this work we have presented the first results of a study of BAL QSOs
(at low and high redshift), based on very deep Gemini GMOS integral field
unit (IFU) spectroscopy.
In particular, the results obtained for the nearest BAL IR--QSO Mrk 231
are presented.
These GMOS data are combined with 3D and 1D spectroscopy
(obtained previously at La Palma/WHT, HST/FOS, and KPNO
observatories) and deep HST broad band images of Mrk 231.

The main results and conclusions can be summarised as follows:

\begin{enumerate}

\item
Very deep three-dimensional (3D) spectra and maps clearly show that
the BAL systems {\sc i} and {\sc ii} --mainly in the strong ``absorption lines"
Na ID$\lambda$5889-95 and Ca {\sc ii} K$\lambda$3933-- are extended
(reaching $\sim$1.4--1.6$''$ $\sim$1.2--1.3 kpc, from the nucleus)
and clearly elongated at the position angle (PA) close to the radio jet PA.
Which suggest that the BAL systems {\sc i} and {\sc ii} are ``both" associated
with the radio jet, and supporting the bipolar jet-wind model for some BALs.

\item
For the nuclear region of Mrk 231, the QSO and host-galaxy components
were modelled, using a new technique of decoupling 3D spectra.

From this study, the following main results were found:

\begin{itemize}

\item
In the pure host galaxy spectrum a strong/extreme nuclear starburst
component was clearly observed (for the first time, at optical wavelength),
mainly as a very strong increase in the flux, at the blue region;

\item
The BAL system {\sc i}  is observed in the spectrum of the host galaxy, i.e.:
confirming their extended morphology;

\item
In the clean/pure QSO emission spectrum, only broad lines were detected.

\item
3D GMOS individual spectra (specially the IR Ca {\sc ii} triplet and [O {\sc ii}]
 $\lambda$3727) and maps confirm
the presence of an extreme young nuclear starburst component. Which was
detected mainly in a toroid or ring at r $=$ 0.3$'' \sim$ 200 pc.

\item
The nuclear starburst plus the bubbles are cleaning the nuclear dust,
specially  in the south region. This area is coincident with the region where
we previously suggested that the galactic wind --with super bubble/shells--
is in the blowout phase.

\end{itemize}

\item
On the other hand, the 3D spectra of the knots of the multiple expanding
nuclear bubbles were analysed. In particular, we found,

\begin{itemize}

\item
Strong multiple emission line systems (with LINER properties) and Wolf Rayet
features in the main knots of the
more external superbubble S1. The kinematics of these knots --and the
internal bubbles-- suggest that these knots
are probably associated with a main area of rupture of the supergiant bubble
(at the south--west region).

\item
In the more internal superbubble S4 and close to the very nucleus
(for r $<$ 0.7$" \sim$ 0.6 kpc), two similar narrow emission
line systems were detected, with strong [S {\sc ii}] and [O {\sc i}] emission
and $\Delta$V $\sim$ --200 km s$^{-1}$. These results suggest that for the
nuclear region an important part of the NLR--emission is generated by the
OF process (and the associated low velocity ionizing shocks).

\end{itemize}

\end{enumerate}

The composite nature of the BAL systems of Mrk 231  is discussed.
In addition, a {\it composite hyper--wind scenario} (already proposed
for BALs) is suggested for the origin of giant Ly$\alpha$ blobs, at high
redshift. The importance of study the end phases of Mrk 231, elliptical
galaxies and QSOs (i.e., galaxy remnants) is briefly discussed.

\section*{Acknowledgments}

Based mainly on observations obtained at the Gemini Observatory, which is operated
by AURA under cooperative agreement with the NSF-USA on behalf of the Gemini
partnership: NSF-USA, PPARC-UK, NRC-Canada, CONICYT-Chile,
ARC-Australia, CNPq-Brazil and CONICET-Argentina.
This research was based also on observations of La Palma and KPNO
observatories, and archive data of the NASA and ESA satellite HST
(obtained from the archive of ESO Garching and STScI--Baltimore).
The authors thank  P. Candia, H. Dottori,
for discussions and assistance.
We specially thank to the Director of Gemini North Observatory and to
the Manager of the Gemini Argentina Office, J. Roy and H. Levato, for
the help in the early schedule (and the change of target) of the Gemini
GMOS observations, of Mrk 231.
Finally, we wish to thank the referee for constructive and
valuable comments, which helped to improve the content
and presentation of the paper.

\clearpage

\begin{table}
\footnotesize \caption{Journal of  observations of Mrk\,231}
\label{tobser}

\begin{tabular}{llllcl}
\hline
\hline
Object & Date & Telescope/ & Spectral region &Expos.\ time   & Comments \\
       &      & instrument &                 & [s]           & \\
\hline

       &      &            &                 &               & \\

       &      &            &                 &               & \\

Gem Data&     &            &                 &               & \\

(North)&      &            &                 &               & \\

Mrk 231& 2005 Apr 06& 8.1\,m Gemini/GMOS-IFU&R831, $\lambda\lambda$5750--7850 \AA&900$\times$1&
Nucleus, $\langle$FWHM$\rangle$= 0\farcs7-0\farcs8 \\

Mrk 231& 2005 Apr 06& 8.1\,m Gemini/GMOS-IFU&R831, $\lambda\lambda$5750--7850 \AA&900$\times$1&
Arc, $\langle$FWHM$\rangle$= 0\farcs7-0\farcs8 \\

Mrk 231& 2005 Apr 30& 8.1\,m Gemini/GMOS-IFU&B600, $\lambda\lambda$3420--6200 \AA&1800$\times$2&
Nucleus, $\langle$FWHM$\rangle$= 0\farcs4-0\farcs5 \\

Mrk 231& 2005 Apr 30& 8.1\,m Gemini/GMOS-IFU&B600, $\lambda\lambda$3420--6200 \AA&1500$\times$2&
Arc, $\langle$FWHM$\rangle$= 0\farcs4-0\farcs5 \\

Mrk 231& 2005 Apr 30& 8.1\,m Gemini/GMOS-IFU&B600, $\lambda\lambda$4550--7400 \AA&1200$\times$2&
Nucleus, $\langle$FWHM$\rangle$= 0\farcs4-0\farcs5 \\

Mrk 231& 2005 Apr 30& 8.1\,m Gemini/GMOS-IFU&B600, $\lambda\lambda$4550--7400 \AA&900$\times$2&
Arc, $\langle$FWHM$\rangle$= 0\farcs4-0\farcs5 \\

Mrk 231& 2005 Apr 30& 8.1\,m Gemini/GMOS-IFU&R831, $\lambda\lambda$7750--9850 \AA&900$\times$1&
Nucleus, $\langle$FWHM$\rangle$= 0\farcs4-0\farcs5 \\

       &    &            &                           &         & \\

HST Data&   &            &                           &         & \\
(Archive)&  &            &                           &         & \\

Mrk 231& 1995 Oct 23&{\itshape HST\/}/WFPC2& F439W, $\lambda\lambda$4283/464 \AA & 2226&
$\langle$FWHM$\rangle$ = 0\farcs1\\
Mrk 231& 1995 Oct 23 &{\itshape HST\/}/WFPC2& F814W, $\lambda \lambda$8203/1758 \AA & 712&
   $''$\\

Mrk 231& 2003 Mar 17&{\itshape HST\/}/ACS  & F330W, $\lambda \lambda$3354/588 \AA & 1140&
$\langle$FWHM$\rangle$ = 0\farcs1\\

Mrk 231& 2003 Sep 09 &{\itshape HST\/}/NICMOS  & F160W, $\lambda \lambda$1.60/0.40 $\mu$m& 640&
$\langle$FWHM$\rangle$ = 0\farcs22\\

       &    &            &                           &         & \\
Mrk 231& 1992 Nov 27&{\itshape HST\/}/FOS& G190H, $\lambda\lambda$1275--2320 \AA & 5760& \\
Mrk 231& 1992 Nov 27 &{\itshape HST\/}/FOS& G270H, $\lambda \lambda$2225--3295 \AA & 2880& \\
Mrk 231& 1996 Nov 21 &{\itshape HST\/}/FOS& G160L, $\lambda \lambda$1150--2300 \AA &  770& \\

       &    &            &                           &         & \\

WHT Data&   &            &                           &         & \\

La Palma&   &            &                           &         & \\

Mrk 231& 2001 Apr 12& 4.2\,m WHT/INTEGRAL&$\lambda\lambda$6200--7600 \AA&1800$\times$3&
$\langle$FWHM$\rangle$ = 1\farcs0 \\

       &    &            &                           &         & \\

NOT Data&   &            &                           &         & \\

La Palma&   &            &                           &         & \\

Mrk 231& 1991 May 11& 2.5\,m NOT   &V                          &1200$\times$3&
$\langle$FWHM$\rangle$ = 0\farcs7 \\

       &    &            &                           &         & \\

KPNO Data&  &            &                           &         & \\

       &    &            &                           &         & \\

Mrk 231& 1991 Feb 15& 2.15\,m KPNO/GoldCam&$\lambda\lambda$3350--5200 \AA&900$\times$2&
PA = 90$^{\circ}$, slit width = 1\farcs5 \\
       &            & 2.15\,m KPNO/GoldCam&$\lambda\lambda$5100--7100 \AA&900$\times$2&
$''$ \\

       &    &            &                           &         & \\
       &    &            &                           &         & \\

\hline

\end{tabular}

\end{table}

\clearpage

\begin{table}
\footnotesize \caption{Main properties of the circumnuclear knots,
associated with the supershells (in Mrk 231)}
\label{tknots}

\begin{tabular}{lrrcclrl}
\hline
\hline
Knots$^a$&$\Delta\alpha$$^b$&$\Delta\delta$&B$_{F439W}^a$&I$_{F814W}^a$&B-I
& R$_{eff}^a$ & Shells \\
         & [$''$]           & [$''$]       & [mag]       & [mag]       &
& [pc]        &    \\
\hline

  &     &     &       &       &      &      &    \\
1 & 1.10& 1.68& 22.23 & 21.12 & 1.11 &    66&  S1\\
2 &-1.52&-0.90& 23.27 & 20.86 & 2.41 & $<$20&  S2\\
3 & 1.10&-0.16& 22.38 & 20.86 & 1.52 &    62&  S3\\
4 & 2.12&-1.45& 23.98 & 22.60 & 1.38 &    42&  S1\\
5 & 1.70&-2.97& 21.31 & 20.50 & 0.81 & $<$20&  S1\\
6 & 1.56&-2.74& 22.97 & 22.28 & 0.69 &    71&  S1\\
7 & 1.01&-3.70& 22.69 & 21.47 & 1.22 &    38&  S1\\
11&-0.46&-3.66& 21.17 & 20.66 & 0.51 &   136&  S1\\
12&-0.69&-3.61& 21.46 & 20.70 & 0.76 &    31&  S1\\
14&-1.20&-3.15& 21.79 & 20.33 & 1.46 &    40&  S1\\
29&-2.30& 1.82& 23.26 & 21.06 & 2.20 &    46&  S1\\
  &     &     &       &       &      &      &    \\

\hline

\end{tabular}

Notes:\\
$^a$: From Surace et al. (1998).\\
$^b$: The offset positions of the knots [$\Delta\alpha$ (RA),$\Delta\delta$
(DEC)] are given from the nucleus position (as 0,0).\\

\end{table}

\clearpage

\begin{table}
\footnotesize \caption{Emission Lines of the main knots of the
shell--S1 (located outside of the south west region)}
\label{flux3ds1a}
\begin{tabular}{llccccc}
\hline
\hline

Lines&Component&   &      &Fluxes$^{a}$& &         \\
     &     &Knot 1 &Knot 4&Knot 5&Knot 6 &Knot 7   \\
     &     &N-B043+R094&N-B022+R021&A-B043+R058&A-B044+R095&A-B140+R160\\

\hline

  &         &      &      &      &      &        \\

H$\beta\lambda4861$
  & MC-EMI  &(0.10)&(0.08)& 0.11 & 0.09 & 0.16 \\
  & OF-EB1  &(0.01)& ---  &(0.01)& 0.04 & ---   \\
  &         &      &      &      &      &       \\

[O{\sc iii}]$\lambda5007$
  & MC-EMI  & 0.05 &(0.10)& 0.17 & 0.22 & 0.15 \\
  & OF-EB1  & 0.03 & ---  & 0.03 & 0.08 & --- \\
  &         &      &      &      &      &       \\

[O {\sc i}]$\lambda6300$
  & MC-EMI  & 0.15 & 0.04 & 0.11 & 0.14 & 0.05  \\
  & OF-EB1  & 0.02 & 0.03 & ---  & ---  & ---   \\
  & OF-ER1  & ---  & ---  & ---  & 0.16 & ---    \\
  &         &      &      &      &      &        \\

H$\alpha\lambda6563$
  & MC-EMI  & 0.85 & 0.22 & 0.39 & 0.35 & 0.54  \\
  & OF-EB1  & 0.06 & 0.05 & 0.04 & 0.11 & 0.06  \\
  & OF-ER1  & 0.03 & 0.04 & 0.05 & 0.09 & 0.02  \\
  &         &      &      &      &      &       \\

[N {\sc ii}]$\lambda6583$
  & MC-EMI  & 0.97 & 0.23 & 0.48 & 0.48 & 0.38  \\
  & OF-EB1  & 0.09 & 0.15 & 0.05 & 0.12 & 0.08  \\
  & OF-ER1  & 0.05 & 0.06 & 0.04 & 0.12 & 0.03  \\
  &         &      &      &      &      &       \\

[S {\sc ii}]$\lambda6717$
  & MC-EMI  & 0.30 & 0.14 & 0.20 & 0.15 & 0.15  \\
  & OF-EB1  & 0.02 & 0.10 & 0.08 & 0.06 & 0.04  \\
  &         &      &      &      &      &        \\

[S {\sc ii}]$\lambda6731$
  & MC-EMI  & 0.28 & 0.11 & 0.15 & 0.10 & 0.13  \\
  & OF-EB1  & 0.03 & 0.09 & 0.05 & 0.06 & 0.04  \\
  &         &      &      &      &      &       \\

H$\alpha$/H$\beta$
  & MC-EMI  &(8.8) &(2.8) & 3.5  & 3.9  & 3.4      \\
  & OF-EB1  &(6.0) & ---  & 4.0  & 2.8  & ---   \\
  &         &      &      &      &      &       \\

FWHM H$\alpha$[km/s] 
  & MC-EMI  & 195 & 135   & 115  & 137  & 120   \\
  & OF-EB1  & 120 & 130   & 110  & 135  & 113   \\
  &         &     &       &      &      &       \\

\hline

\end{tabular}

\noindent
$^{a}$: the fluxes are given in units of 10$^{-16}$ erg cm$^{-2}$ s$^{-1}$
(from GMOS/IFU-B600).\\
Line 2: N-Bmmm+Rnnn mean B600 spectrum at Nucleus frame, Blue mmm and
Red nnn number of the lenss.\\
Column 2: emission line components, where
MC-EMI,  OF-EB1, and OF-ER1 are the \\
main-component at Z $=$ 0.04250 (cz $=$ 12750 km/s); \\
OF blue component 1 at Z $=$ [0.04140, 0.04210] (cz $=$ 12470, 12620 km/s), 
$\Delta$V $=$ [--150 -- 300] km s$^{-1}$; \\
OF blue component 3 at Z $=$ 0.03920 (11749 km/s),
$\Delta$V $=$ --905 km s$^{-1}$); \\
OF red  component 1 at Z $=$ [0.04298, 0.04333] (cz $=$ 12895, 12998 km/s), 
$\Delta$V $=$ [+150 , +250] km s$^{-1}$, respectively.\\
The values between parentheses are data with low S/N.\\

\end{table}

\clearpage

\begin{table}
\footnotesize \caption{Emission Lines of the strong knots of the
shell--S1 and the Region SW1 (located at the south-west region)}
\label{flux3ds1b}
\begin{tabular}{llcccccc}
\hline
\hline

Lines&Component&       &Fluxes$^{a}$&        &          &   & \\
     &      &Knot 11   &Knot 12   &Knot 14W  &Knot 14E  &---&Region SW1 \\
     &      &A-B311+R342 &A-B341+R392 &A-B408+R445 &A-B359+R407 &   &A-B480+R482 \\

\hline

     &      &          &          &          &          &   &  \\

WR bump$\lambda4650$
  &MC-EMI$^*$& ---     & ---      & (0.19)   & 0.32     &   & --- \\
     &      &          &          &          &          &   &  \\

H$\beta\lambda4861$
  & MC-EMI$^*$& 0.10   & 0.07     & 0.09     & 0.11     &   & 0.06 \\
  & OF-EB2$^*$& ---    & ---      & 0.05     & 0.06     &   & --- \\
  & OF-EB3$^*$& ---    & ---      & ---      & ---      &   &(0.07) \\
  & OF-ER1$^*$& ---    & 0.11     & ---      & 0.07     &   & ---  \\
  &         &          &          &          &          &   &      \\

[O{\sc iii}]$\lambda5007$
  & MC-EMI$^*$& 0.15   & 0.24     & 0.58     & 0.42     &   & 0.18 \\
  & OF-EB2$^*$& 0.08   & 0.06     & 0.08     & 0.07     &   & ---  \\
  & OF-EB3$^*$& ---    & ---      & ---      & ---      &   & 0.40 \\
  & OF-ER1$^*$& 0.10   & 0.10     & ---      & 0.08     &   & ---  \\
  &         &          &          &          &          &   &      \\

[N {\sc i}]$\lambda5198$
  &MC-EMI$^*$& ---     & ---      & ---      & (0.11)   &   & ---  \\
  &         &          &          &          &          &   &      \\

[O {\sc i}]$\lambda6300$
  & MC-EMI$^*$& 0.06   & 0.02     & 0.03     & 0.05     &   & 0.03 \\
  & OF-EB2$^*$& ---    & ---      & 0.04     & 0.03     &   & ---  \\
  & OF-EB3$^*$& ---    & ---      & ---      & ---      &   & 0.03 \\
  & OF-ER1$^*$& 0.04   & 0.03     & 0.04     & 0.03     &   & ---  \\
  &         &          &          &          &          &   &      \\

H$\alpha\lambda6563$
  & MC-EMI$^*$& 0.66   & 0.49     & 0.23     & 0.40     &   & 0.17 \\
  & OF-EB2$^*$& 0.09   & 0.12     & 0.37     & 0.35     &   & ---  \\
  & OF-EB3$^*$& ---    & ---      & ---      & ---      &   & 0.13 \\
  & OF-ER1$^*$& 0.01   & 0.03     & 0.07     & 0.08     &   & ---  \\
  &         &          &          &          &          &   &      \\

[N {\sc ii}]$\lambda6583$
  & MC-EMI$^*$& 0.43   & 0.33     & 0.29     & 0.48     &   & 0.20 \\
  & OF-EB2$^*$& 0.17   & 0.12     & 0.25     & 0.28     &   & ---  \\
  & OF-EB3$^*$& ---    & ---      & ---      & ---      &   & 0.14 \\
  & OF-ER1$^*$& 0.12   & 0.10     & 0.06     & 0.05     &   & ---  \\
  &         &          &          &          &          &   &      \\

[S {\sc ii}]$\lambda6717$
  & MC-EMI$^*$& 0.18   & 0.20     & 0.24     & 0.23     &   & 0.13 \\
  & OF-EB2$^*$& 0.03   & ---      &(0.15)    & 0.06     &   & ---  \\
  & OF-EB3$^*$& ---    & ---      & ---      & ---      &   & 0.12 \\
  &         &          &          &          &          &   &      \\

[S {\sc ii}]$\lambda6731$
  & MC-EMI$^*$& 0.15   & 0.18     & 0.18     & 0.14     &   & 0.11 \\
  & OF-EB2$^*$& ---    & ---      &(0.15)    & 0.06     &   & ---  \\
  & OF-EB3$^*$& ---    & ---      & ---      & ---      &   & 0.11 \\
  &         &          &          &          &          &   &      \\

H$\alpha$/H$\beta$
  & MC-EMI$^*$& 6.6    & 7.0      & 2.7      & 3.6      &   & 2.8  \\
  & OF-EB2$^*$& ---    & ---      & 7.4      & 5.8      &   & ---  \\
  & OF-EB3$^*$& ---    & ---      & ---      & ---      &   &(2.0)  \\
  &         &          &          &          &          &   &      \\

FWHM H$\alpha$ [km/s]
  & MC-EMI$^*$& 150    & 140      & 120      & 165      &   & 198  \\
  & OF-EB2$^*$& 115    & 149      & 125      & 115      &   & ---  \\
  & OF-EB3$^*$& ---    & ---      & ---      & ---      &   & 165  \\
  &         &          &          &          &          &   &      \\

\hline

\end{tabular}

\noindent
$^{a}$: the fluxes are given in units of 10$^{-16}$ erg cm$^{-2}$ s$^{-1}$
(from GMOS/IFU-B600).\\
Line 2: N-Bmmm+Rnnn mean B600 spectrum at Nucleus frame, Blue mmm and
Red nnn number of the lenss.\\
Column 2: emission line components, where
MC-EMI$^*$,  OF-EB2$^*$, OF-EB3$^*$, OF-ER1$^*$ are the \\
main-component at SW region with Z $=$ 0.04220 (cz $=$ 12600 km/s); \\
OF blue component 2 at Z $=$ 0.04080 (12240 km/s),
$\Delta$V $=$ --400 km s$^{-1}$); \\
OF blue component 3 at Z $=$ 0.03920 (11749 km/s),
$\Delta$V $=$ --905 km s$^{-1}$); \\
OF red  component 1 at south west region,
at Z $\sim$ 0.04280 (cz $=$ 12840 km/s),
$\Delta$V $\sim$ +200 km s$^{-1}$.\\
The values between parentheses are data with low S/N.\\

\end{table}

\clearpage

\begin{table}
\footnotesize \caption{Emission Lines of the main knots of the
shells S2 (located to the south west region, of the nucleus)}
\label{flux3ds2}
\begin{tabular}{llccccccc}
\hline
\hline

Lines&Compon&    &Fluxes$^{a}$&        &        &        &        &        \\
     &         &Knot S2a&Knot S2b&Knot S2c&Knot S2d&Knot S2e&Knot S2f&Knot S2g\\
     &         &N-B419+R434&N-B382+R418&N-B331+R369&N-B380+R320&N-B122+R131&N-B039+R090
&N-B159+R208\\
     &         &[1.4$''$W,0.8$''$S]&[1.2$''$W,0.9$''$S]&[0.9$''$W,1.1$''$S]&[0.5$''$W,1.3$''$S]
&[0.7$''$E,1.4$''$S]&[1.0$''$E,0.4$''$N]&[0.2$''$E,1.1$''$N] \\
  &            &        &        &        &        &        &        &        \\

\hline

  &            &        &        &        &        &        &        &        \\

H$\beta\lambda4861$
  & MC-EMI     & ----   & ---    & ---    & ---    & 0.07   & 0.21   & 0.41   \\
  & MC-EMI$^*$ & 0.07   & 0.09   & 0.08   & 0.11   & ---    & ---    & ---    \\
  & OF-EB2$^*$ & 0.04   & ---    & ---    & ---    & ---    & ---    & ----   \\
  &            &        &        &        &        &        &        &        \\

[O{\sc iii}]$\lambda5007$
  & MC-EMI     & ----   & ---    & ---    & ---    & 0.20   & 0.20   & 0.40   \\
  & MC-EMI$^*$ & 0.18   & 0.18   & 0.08   & 0.11   & ---    & ---    & ---    \\
  & OF-EB2$^*$ & 0.14   & ---    & ---    & ---    & ---    & ---    & ---    \\
  &            &        &        &        &        &        &        &        \\

[O {\sc i}]$\lambda6300$
  & MC-EMI     & ----   & ---    & ---    & ---    & 0.05   & 0.11   & 0.11   \\
  & MC-EMI$^*$ & 0.12   & 0.10   & 0.14   & 0.05   & ---    & ---    & ---     \\
  & OF-EB2$^*$ & 0.03   & 0.05   & 0.02   & 0.02   & ---    & ---    & ---    \\
  &            &        &        &        &        &        &        &        \\

H$\alpha\lambda6563$
  & MC-EMI     & ----   & ---    & ---    & ---    & 0.21   & 1.66   &  1.70  \\
  & MC-EMI$^*$ & 0.30   & 0.36   & 0.45   & 0.40   & ---    & ---    &  ---  \\
  & OF-EB1     & ----   & ---    & ---    & ---    & ---    & ---    &  0.10  \\
  & OF-EB2$^*$ & 0.10   & 0.13   & 0.07   & 0.07   & ---    & ---    &  ---  \\
  & OF-ER1$^*$ & 0.02   & 0.02   & 0.11   & 0.04   & ---    & ---    &  ---  \\
  &            &        &        &        &        &        &        &       \\

[N {\sc ii}]$\lambda6583$
  & MC-EMI     & ----   & ---    & ---    & ---    & 0.41   & 1.22   &  1.27  \\
  & MC-EMI$^*$ & 0.40   & 0.45   & 0.44   & 0.44   & ---    & ---    &  ---  \\
  & OF-EB1     & ----   & ---    & ---    & ---    & 0.06   & ---    &  0.07   \\
  & OF-EB2$^*$ & 0.10   & 0.15   & 0.06   & 0.11   & ---    & ---    &  ---  \\
  & OF-ER1$^*$ & 0.04   & 0.02   & 0.05   & 0.06   & ---    & ---    &  ---  \\
  &            &        &        &        &        &        & ---    &  ---  \\

[S {\sc ii}]$\lambda6717$
  & MC-EMI     & ----   & ---    & ---    & ---    & 0.18   & 0.40   &  0.70  \\
  & MC-EMI$^*$ & 0.22   & 0.23   & 0.10   & 0.13   & ---    & ---    &  ---   \\
  & OF-EB1     & ----   & ---    & ---    & ---    & ---    & ---    &  0.20  \\
  & OF-EB2$^*$ & 0.02   & 0.04   & 0.04   & 0.06   & ---    & ---    &  ---  \\
  & OF-ER1$^*$ & ---    & ---    & 0.04   & 0.02   & ---    & ---    &  ---  \\
  &            &        &        &        &        &        &        &       \\

[S {\sc ii}]$\lambda6731$
  & MC-EMI     & ----   & ---    & ---    & ---    & 0.10   & 0.34   &  0.60 \\
  & MC-EMI$^*$ & 0.16   & 0.24   & 0.17   & 0.12   & ---    & ---    &  --- \\
  & OF-EB1     & ----   & ---    & ---    & ---    & ---    & ---    &  0.16 \\
  & OF-EB2$^*$ & 0.03   & 0.06   & 0.04   & 0.01   & ---    & ---    &  --- \\
  & OF-ER1$^*$ & ---    & ---    & 0.06   & 0.01   & ---    & ---    &  --- \\
  &            &        &        &        &        &        &        &      \\
  &            &        &        &        &        &        &        &      \\

H$\alpha$/H$\beta$
  & MC-EMI     & ----   & ---    & ---    & ---    & 3.0    & 7.9    & 4.2  \\
  & MC-EMI$^*$ & 4.3    & 8.0    & 5.6    & 3.6    & ---    & ---    & ---  \\
  & OF-EB1     & ----   & ---    & ---    & ---    & ---    & ---    & ---  \\
  & OF-EB2$^*$ & 2.5    & ---    & ---    & ---    & ---    & ---    & ---  \\
  &            &        &        &        &        &        &        &     \\

FWHM 
  & MC-EMI     & ----   & ---    & ---    & ---    &  130   & 135    & 185 \\
H$\alpha$
  & MC-EMI$^*$ & 120    & 150    & 215    & 220    &  ---   & ---    & --- \\
  & OF-EB1     & ----   & ---    & ---    & ---    &  ---   & ---    & ---    \\
  & OF-EB2$^*$ & 120    & 165    & 150    & 170    &  ---   & ---    & --- \\

  &            &        &        &        &        &        &        &     \\

\hline

\end{tabular}

\noindent
$^{a}$: the fluxes are given in units of 10$^{-16}$ erg cm$^{-2}$ s$^{-1}$
(from GMOS/IFU-B600 spectroscopy).\\
Column 2: emission line components, as in Table 4.\\
Line 4: the RA and DEC off set (from the very nucleus, as 0,0)
for each GMOS spectrum, in each knot.\\
The values between parentheses are data with low S/N.\\
The FWHM unit is [km/s].\\

\end{table}

\clearpage

\begin{table}
\footnotesize \caption{Emission Lines of the main knots of the
shells S3 and Region E1}
\label{flux3ds3}
\begin{tabular}{llccccccc}
\hline
\hline

Lines &Compon&     &Fluxes$^{a}$&      &        &        &      &           \\
      &     &Knot S3a &Knot S3b &Knot S3c &Knot S3d&Knot S3e&Knot S3f& Region E1 \\
      &     &N-B085+R115&N-B118+R135&N-B169+R184&N-B318+R335&N-B317+R336&N-B113+R140&N-B065+R088  \\
      &     &[0.9$''$E,0.3$''$S]&[0.7$''$E,0.6$''$S]&[0.4$''$E,0.8$''$S]&[0.7$''$W,0.6$''$S]
&[0.7$''$W,0.4$''$S]&[0.7$''$E,0.4$''$N]&[1.0$''$E,0.0$''$] \\
      &     &         &         &         &        &        &      &           \\

\hline

  &         &         &         &         &        &        &      &   \\

H$\beta\lambda4861$
  & MC-EMI  & 0.51    & 0.20    & 0.09    & ---    & ---    & 0.20 & 0.73   \\
  &MC-EMI$^*$& ---    & ---     & ---     & 0.08   & 0.12   & ---  & ---   \\
  & OF-EB1  & ---     & ---     & ----    & ---    & ---    & 0.09 & 0.10   \\
  &OF-EB2$^*$& ---    & ---     & ---     & 0.06   & ---    & ---  & ---   \\
  &         &         &         &         &        &        &      &        \\

[O{\sc iii}]$\lambda5007$
  & MC-EMI  & 0.25    & 0.10    & 0.10    & ---    & ---    & 0.10 & 0.39  \\
  &MC-EMI$^*$& ---    & ---     & ---     & 0.08   & 0.13   & ---  & ---   \\
  & OF-EB1  & ---     & ---     & ---     & ---    & ---    & 0.07 & 0.09  \\
  &OF-EB2$^*$& ---    & ---     & ---     & 0.07   & ---    & ---  & ---   \\
  &         &         &         &         &        &        &      &       \\

[O {\sc i}]$\lambda6300$
  & MC-EMI  & 0.07    & 0.08    & 0.16    &  ---   & ---    & 0.18 & 0.13  \\
  &MC-EMI$^*$& ---    & ---     & ---     &  0.20  & 0.15   & ---  & ---   \\
  & OF-EB1  & 0.07    & 0.06    & 0.05    &  ---   & ---    & ---  & 0.02  \\
  &OF-EB2$^*$& ---    & ---     & ---     &  0.19  & 0.26   & ---  & ---   \\
  &         &         &         &         &        &        &      &       \\

H$\alpha\lambda6563$
  & MC-EMI  & 4.50    & 1.04    & 0.80    &  ---   & ---    & 1.26 & 6.93   \\
  &MC-EMI$^*$& ---    & ---     & ---     &  0.55  & 0.83   & ---  & ---   \\
  & OF-EB1  & ---     & 0.35    & 0.30    &  ---   & ---    & ---  & ---   \\
  &OF-EB2$^*$& ---    & ---     & ---     &  ---   & ---    & ---  & ---   \\
  &         &         &         &         &        &        &      &        \\

[N {\sc ii}]$\lambda6583$
  & MC-EMI  &  2.70   & 0.68    & 0.68    &  ---   & ---    & 1.17 & 4.42  \\
  &MC-EMI$^*$& ---    & ---     & ---     &  0.20  & 0.31   & ---  & ---   \\
  & OF-EB1  & ---     & 0.11    & 0.10    &  ---   & ---    & ---  & ---    \\
  &OF-EB2$^*$& ---    & ---     & ---     &  ---   & ---    & ---  & ---   \\
  &         &         &         &         &        &        &      &        \\

[S {\sc ii}]$\lambda6717$
  & MC-EMI  & 0.47    & 0.21    & 0.20    &  ---   & ---    & 0.42 & 0.60  \\
  &MC-EMI$^*$& ---    & ---     & ---     &  0.24  & 0.22   & ---  & ---   \\
  & OF-EB1  & ---     & 0.06    & 0.07    &  ---   & ---    & ---  & ---   \\
  &OF-EB2$^*$& ---    & ---     & ---     &  ---   & 0.21   & ---  & ---   \\
  & OF-ER1  & ---     & ---     & 0.04    &  ---   & ---    & ---  & ---   \\
  &         &         &         &         &        &        &      &        \\

[S {\sc ii}]$\lambda6731$
  & MC-EMI  & 0.47    & 0.20    & 0.15    &  ---   & ---    & 0.35 & 0.59   \\
  &MC-EMI$^*$& ---    & ---     & ---     &  0.20  & 0.18   & ---  & ---   \\
  & OF-EB1  & ---     & 0.03    & 0.04    &  ---   & ---    & ---  & ---    \\
  &OF-EB2$^*$& ---    & ---     & ---     &  ---   & 0.11   & ---  & ---   \\
  & OF-ER1  & ---     & ---     & 0.05    &  ---   & ---    & ---  & ---   \\
  &         &         &         &         &        &        &      &       \\
  &         &         &         &         &        &        &      &       \\

H$\alpha$/H$\beta$
  & MC-EMI  & 8.8     & 5.2     & 8.8     & ---    & ---    & 6.3  & 9.4  \\
  &MC-EMI$^*$&---     & ---     & ---     & 6.9    & 6.8    & ---  & ---   \\
  & OF-EB1  & ---     & ---     & ---     & ---    & ---    & ---  & ---  \\
  &OF-EB2$^*$&---     & ---     & ---     & ---    & ---    & ---  & ---   \\
  &         &         &         &         &        &        &      &       \\

FWHM 
  & MC-EMI  &  113    & 105     & 166     & ---    & ---    & 150  & 120 \\
H$\alpha$
  &MC-EMI$^*$& ---    & ---     & ---     & 115    & 128    & ---  & ---   \\
  & OF-EB1  &  ---    & 180     & 183     & ---    & ---    & ---  & ---  \\
  &OF-EB2$^*$& ---    & ---     & ---     & ---    & ---    & ---  & ---   \\

  &         &         &         &         &        &        &      &       \\

\hline

\end{tabular}

\noindent
$^{a}$: the fluxes are given in units of 10$^{-16}$ erg cm$^{-2}$ s$^{-1}$
(from GMOS/IFU-B600 spectroscopy).\\
Column 2: emission line components, as in Tables 3 and 4.\\
Line 4: the RA and DEC off set (from the very nucleus, as 0,0)
for each GMOS spectrum, in each knot.\\
The values between parentheses are data with low S/N.\\

\end{table}

\clearpage

\begin{table}
\footnotesize \caption{Emission Lines of the main knots of the
shell S4}
\label{flux3ds4}
\begin{tabular}{llcccccc}
\hline
\hline

Lines&Component&    &Fluxes$^{a}$&    &        &        &   \\
     &     &Knot S4a&Knot S4b&Knot S4c&Knot S4d&Knot S4e&Knot S4f \\
     &     &B163+R190&B188+R212&B267+R286&B288+R313&B285+R315&B167+R186\\
     &     &[0.4$''$E,0.4$''$N]&[0.2$''$E,0.3$''$N]&[0.4$''$W,0.4$''$S]&[0.5$''$W,0.1$''$N]
&[0.5$''$W,0.3$''$S]&[0.4$''$E,0.4$''$S]          \\
  &        &       &       &       &      &       &          \\

\hline

  &        &       &       &       &      &       &          \\

H$\beta\lambda4861$
  & MC-EMI & 0.16  & 0.30  & 0.08  & 0.11 & ---   &  0.13   \\
  &MC-EMI$^*$& --- & ---   & ---   & ---  & 0.11  &  ---    \\
  & OF-EB1 & ---   & ---   & 0.07  & 0.08 & ---   &  ---    \\
  &OF-EB2$^*$& --- & ---   & ---   & ---  & 0.09  &  ---    \\
  &        &       &       &       &      &       &        \\

[O{\sc iii}]$\lambda5007$
  & MC-EMI & 0.09  & 0.17  & 0.19  & 0.15 & ---   &  0.20   \\
  &MC-EMI$^*$& --- & ---   & ---   & ---  & 0.14  &  ---    \\
  & OF-EB1 & ---   & ---   & 0.12  & 0.06 & ---   &  ---    \\
  &OF-EB2$^*$& --- & ---   & ---   & ---  & 0.11  &  ---    \\
  &        &       &       &       &      &       &         \\

[O {\sc i}]$\lambda6300$
  & MC-EMI & 0.33  & 0.43  & 0.29  & 0.20 & ---   &  0.23  \\
  &MC-EMI$^*$& --- & ---   & ---   & ---  & 0.50  &  ---   \\
  & OF-EB1 & 0.06  & ---   & 0.30  & 0.31 & ---   &  ---   \\
  &OF-EB2$^*$& --- & ---   & ---   & ---  & 0.55  &  ---   \\
  &        &       &       &       &      &       &         \\

H$\alpha\lambda6563$
  & MC-EMI & 0.90  & 1.05  & 0.98  & 0.63 & ---   &  0.36  \\
  &MC-EMI$^*$& --- & ---   & ---   & ---  & 0.48  &  ---   \\
  & OF-EB1 & ---   & ---   & ---   & ---  & ---   &  ---   \\
  &OF-EB2$^*$& --- & ---   & ---   & ---  & ---   &  ---   \\
  &        &       &       &       &      &       &        \\

[N {\sc ii}]$\lambda6583$
  & MC-EMI & 0.25  & 0.45  &(0.40) &(---) & ---   &  0.15  \\
  &MC-EMI$^*$& --- & ---   & ---   & ---  &(---)  &  ---   \\
  & OF-EB1 & ---   & ---   & ---   & ---  & ---   &  ---   \\
  &OF-EB2$^*$& --- & ---   & ---   & ---  & ---   &  ---   \\
  &        &       &       &       &      &       &        \\

[S {\sc ii}]$\lambda6717$
  & MC-EMI & 0.31  & 0.45  & 0.50  & 0.47 & ---   &  0.56 \\
  &MC-EMI$^*$& --- & ---   & ---   & ---  & 0.33  &  ---  \\
  & OF-EB1 & 0.22  & 0.37  & 0.40  & 0.48 & ---   &  0.42  \\
  &OF-EB2$^*$& --- & ---   & ---   & ---  & 0.60  &  ---   \\
  &        &       &       &       &      &       &         \\

[S {\sc ii}]$\lambda6731$
  & MC-EMI & 0.35  & 0.48  & 0.53  & 0.45 & ---   &  0.48 \\
  &MC-EMI$^*$& --- & ---   & ---   & ---  & 0.45  &  ---  \\
  & OF-EB1 & 0.20  & 0.29  & 0.45  & 0.40 & ---   &  0.40 \\
  &OF-EB2$^*$& --- & ---   & ---   & ---  & 0.68  &  ---  \\
  &        &       &       &       &      &       &         \\

H$\alpha$/H$\beta$

  & MC-EMI & 5.6   & 3.5   & 8.9   & 5.7  & ---   &  2.8    \\
  &MC-EMI$^*$&---  & ---   & ---   & ---  & 4.4   &  ---   \\
  &        &       &       &       &      &       &         \\

FWHM H$\alpha$ [km/s]
  & MC-EMI & 150   &  95   &  98   &  78  & ---   &   85    \\
  &MC-EMI$^*$& --- & ---   & ---   & ---  &  80   &  ---   \\
  &        &       &       &       &      &       &         \\

\hline

\end{tabular}

\noindent
$^{a}$: the fluxes are given in units of 10$^{-16}$ erg cm$^{-2}$ s$^{-1}$
(from GMOS/IFU-B600 spectroscopy).\\
Column 2: emission line components, as in Table 3 and 4.\\
Line 4: the RA and DEC off set (from the very nucleus, as 0,0)
for each GMOS spectrum, in each knot.\\
The values between parentheses are data with low S/N.\\

\end{table}

\clearpage

\begin{table}
\footnotesize \caption{Emission Line Ratios of the main knots of the
shell S1 and the region SW1}
\label{elrs1}
\begin{tabular}{lcccccc}
\hline \hline

               &       &       &       &       &      &        \\
Regions        &Compon& lg[O{\sc iii}]/H$\beta$$^{a}$&lg[O{\sc i}]/H$\alpha^{a}$ 
& lg[N{\sc ii}]/H$\alpha$$^{a}$  & lg[S{\sc ii}s]/H$\alpha$$^{a}$ & Spectral Type \\
\hline
               &       &       &       &       &       &       \\

{\it Knot 1}   &       &       &       &       &       &       \\
               &MC-EMI &(-0.10)& -0.75 &  0.06 & -0.17 & LINER     \\
               &OF-EB1 & (---) & -0.48 &  0.18 & -0.08 & LINER    \\

{\it Knot 4}   &       &       &       &       &       &       \\
               &MC-EMI & (0.00)& -0.74 &  0.02 &  0.06 & LINER     \\
               &OF-EB1 &  ---  & -0.22 &  0.48 &  0.58 & LINER    \\

{\it Knot 5}   &       &       &       &       &       &       \\
               &MC-EMI &  0.19 & -0.55 &  0.09 & -0.05 & LINER     \\
               &OF-EB1 & (---) & ---   &  0.10 &  0.51 & LINER    \\

{\it Knot 6}   &       &       &       &       &       &       \\
               &MC-EMI &  0.39 & -0.40 &  0.14 & -0.15 & LINER     \\
               &OF-EB1 &  0.30 & ---   &  0.04 &  0.04 & LINER    \\

{\it Knot 7}   &       &       &       &       &       &       \\
               &MC-EMI & -0.03 & -1.03 & -0.15 & -0.28 & LINER + H {\sc ii}     \\
               &OF-EB1 &  ---  &  ---  &  0.13 &  0.13 & LINER   \\

{\it Knot 11}  &          &     &       &       &       &       \\
               &MC-EMI$^*$& 0.18& -1.04 & -0.19 & -0.30 & LINER     \\
               &OF-EB2$^*$& --- & ---   &  0.28 & ---   & ---     \\

{\it Knot 12}  &          &     &       &       &       &       \\
               &MC-EMI$^*$& 0.53& -1.39 & -0.17 & -0.11 & LINER     \\
               &OF-EB2$^*$& --- & ---   &  0.00 & ---   & ---     \\

{\it Knot 14W} &          &     &       &       &       &       \\
               &MC-EMI$^*$& 0.76& -0.88 &  0.10 &  0.26 & LINER     \\
               &OF-EB2$^*$& 0.20& -0.97 & -0.17 &(-0.09)& ---     \\

{\it Knot 14E} &       &        &       &       &       &       \\
               &MC-EMI$^*$& 0.58& -0.90 &  0.08 & -0.03 & LINER     \\
               &OF-EB2$^*$& 0.07& -1.12 & -0.10 &(-0.46)& ---     \\

               &          &     &       &       &       &        \\
               &          &     &       &       &       &        \\
{\it Region SW1}&         &     &       &       &       &       \\
               &MC-EMI$^*$& 0.48& -0.75 &  0.07 &  0.15 & LINER     \\
               &OF-EB3$^*$& 0.56& -0.52 &  0.15 & (0.36)& ---     \\

               &          &     &       &       &       &       \\

\hline

\end{tabular}

\noindent
$^{a}$: [O {\sc iii}]$\lambda5007$; [O {\sc i}]$\lambda6300$; 
[N {\sc ii}]$\lambda6583$; [S {\sc ii}s]$\lambda\lambda$6716+6731.\\
Column 2: emission line components (Compon),
as in Table 3.\\
Column 7: spectral type,  using mainly the diagrams
log  log [S {\sc ii}]/H$\alpha$ vs log [O {\sc i}]/H$\alpha$,
log  log [O {\sc iii}]/H$\beta$ vs log [S {\sc ii}]/H$\alpha$, 
log  log [O {\sc iii}]/H$\beta$ vs log [O {\sc i}]/H$\alpha$
(from Heckman et al. 1990: Figure 14; Lipari et al. 2004d).
The values between parentheses are data with low S/N.

\end{table}

\clearpage

\begin{table}
\footnotesize \caption{Emission Line Ratios of the main knots of the
shell S2}
\label{elrs2}
\begin{tabular}{lcccccc}
\hline \hline

               &       &       &       &       &       &       \\
Regions        &Compon& lg[O{\sc iii}]/H$\beta$$^{a}$&lg[O{\sc i}]/H$\alpha^{a}$ 
& lg[N{\sc ii}]/H$\alpha$$^{a}$  & lg[S{\sc ii}s]/H$\alpha$$^{a}$ & Spectral Type \\
\hline
               &          &     &       &       &       &       \\

{\it Knot S2a} &          &     &       &       &       &       \\
               &MC-EMI$^*$& 0.41& -0.40 &  0.13 &  0.10 & LINER     \\
               &OF-EB2$^*$& 0.37& -0.43 &  0.10 & -0.20 & LINER   \\

{\it Knot S2b} &          &     &       &       &       &       \\
               &MC-EMI$^*$& 0.30& -0.56 &  0.10 &  0.02 & LINER     \\
               &OF-EB2$^*$& --- & -0.41 &  0.06 & -0.11 & LINER   \\

{\it Knot S2c} &          &     &       &       &       &       \\
               &MC-EMI$^*$& 0.43& -0.51 & -0.01 & -0.21 & LINER     \\
               &OF-EB2$^*$& --- & -0.54 & -0.06 &  0.06 & LINER   \\

{\it Knot S2d} &          &     &       &       &       &       \\
               &MC-EMI$^*$& 0.00& -0.90 &  0.04 & -0.20 & LINER     \\
               &OF-EB2$^*$& --- & -0.54 &  0.20 &  0.10 & LINER   \\

{\it Knot S2e} &          &     &       &       &       &       \\
               &MC-EMI    & 0.46& -0.62 &  0.29 &  0.12 & LINER     \\
               &OF-EB1    & --- & ---   &  ---  &  ---  & ---     \\

{\it Knot S2f} &          &     &       &       &       &       \\
               &MC-EMI    &-0.02& -1.18 & -0.13 & -0.35 & LINER + H {\sc ii}   \\
               &OF-EB1    & --- & ---   &  ---  &  ---  & ---     \\

{\it Knot S2g} &          &     &       &       &       &       \\
               &MC-EMI    &-0.01& -1.19 & -0.12 & -0.11 & LINER + H {\sc ii}    \\
               &OF-EB1    & --- & ---   & -0.15 & ---   & ---     \\
               &          &     &       &       &       &       \\

\hline

\end{tabular}

\noindent
$^{a}$: [O {\sc iii}]$\lambda5007$; [O {\sc i}]$\lambda6300$; 
[N {\sc ii}]$\lambda6583$; [S {\sc ii}s]$\lambda\lambda$6716+6731.\\
Column 2: emission line components (Compon),
as in Table 3.\\
The values between parentheses are data with low S/N.

\end{table}

\clearpage

\begin{table}
\footnotesize \caption{Emission Line Ratios of the main knots of the
shell S3 and the Region E1}
\label{elrs3}
\begin{tabular}{lcccccc}
\hline \hline

               &       &       &       &       &       &        \\
Regions        &Compon& lg[O{\sc iii}]/H$\beta$$^{a}$&lg[O{\sc i}]/H$\alpha^{a}$ 
& lg[N{\sc ii}]/H$\alpha$$^{a}$  & lg[S{\sc ii}s]/H$\alpha$$^{a}$ & Spectral Type \\
\hline
               &       &       &       &       &       &        \\

{\it Knot S3a} &          &     &       &       &       &       \\
               &MC-EMI    &-0.31& -1.80 & -0.22 & -0.68 & H {\sc ii}\\
               &OF-EB1    & --- & ---   &  ---  &  ---  & ---     \\

{\it Knot S3b} &          &     &       &       &       &       \\
               &MC-EMI    &-0.34& -1.11 & -0.19 & -0.40 & LINER  + H {\sc ii} \\
               &OF-EB1    & --- & -0.77 & -0.50 & -0.59 & LINER    \\

{\it Knot S3c} &          &     &       &       &       &       \\
               &MC-EMI    & 0.18& -0.70 & -0.07 & -0.36 & LINER     \\
               &OF-EB1    & --- & -0.78 & -0.48 & -0.44 & LINER   \\

{\it Knot S3d} &          &     &       &       &       &       \\
               &MC-EMI$^*$& 0.00& -0.44 & -0.44 & -0.10 & LINER     \\
               &OF-EB2$^*$& 0.07& ---   & ---   & ---   & ---     \\

{\it Knot S3e} &          &     &       &       &       &       \\
               &MC-EMI$^*$& 0.03& -0.74 & -0.43 & -0.32 & LINER     \\
               &OF-EB2$^*$& --- & ---   &  ---  & ---   & ---     \\

{\it Knot S3f} &          &     &       &       &       &       \\
               &MC-EMI    &-0.30& -0.85 & -0.03 & -0.21 & LINER     \\
               &OF-EB1    & --- & ---   &  ---  &  ---  & ---     \\

               &       &       &       &       &       &         \\
               &       &       &       &       &       &         \\

{\it Region E1}&       &       &       &       &       &       \\
               &MC-EMI & -0.19 & -1.73 & -0.19 & -0.77 & H {\sc ii}  \\
               &OF-EB1 & -0.05 & ---   & ---   & ---   & ---   \\

               &       &       &       &       &       &        \\

\hline

\end{tabular}

\noindent
$^{a}$: [O {\sc iii}]$\lambda5007$; [O {\sc i}]$\lambda6300$; 
[N {\sc ii}]$\lambda6583$; [S {\sc ii}s]$\lambda\lambda$6716+6731;
[S {\sc ii}]/[S {\sc ii}] $\lambda6716$/$\lambda6731$.\\
Column 2: emission line components (Compon),
as in Table 3.\\
The values between parentheses are data with low S/N.

\end{table}

\clearpage

\begin{table}
\footnotesize \caption{Emission Line Ratios of the main knots of the
shell S4}
\label{elrs4}
\begin{tabular}{lcccccc}
\hline \hline

               &       &       &       &       &       &      \\
Regions        &Compon& lg[O{\sc iii}]/H$\beta$$^{a}$&lg[O{\sc i}]/H$\alpha^{a}$ 
& lg[N{\sc ii}]/H$\alpha$$^{a}$  & lg[S{\sc ii}s]/H$\alpha$$^{a}$ & Spectral Type \\
\hline
               &       &       &       &       &       &       \\

{\it Knot S4a} &          &     &       &       &       &       \\
               &MC-EMI    &-0.25& -0.44 & -0.56 & -0.13 & LINER     \\
               &OF-EB1    & --- & ---   &  ---  &  ---  & ---     \\

{\it Knot S4b} &          &     &       &       &       &       \\
               &MC-EMI    &-0.25& -0.39 & -0.38 & -0.05 & LINER     \\
               &OF-EB1    & --- & ---   & ---   & ---   & ---     \\

{\it Knot S4c} &          &     &       &       &       &       \\
               &MC-EMI    & 0.38& -0.53 &(-0.39)&  0.02 & LINER     \\
               &OF-EB1    & 0.23& ---   & ---   & ---   & ---     \\

{\it Knot S4d} &          &     &       &       &       &       \\
               &MC-EMI    & 0.14& -0.50 & ---   &  0.17 & LINER     \\
               &OF-EB1    &-0.12& ---   & ---   & ---   & ---     \\

{\it Knot S4e} &          &     &       &       &       &       \\
               &MC-EMI$^*$& 0.11&  0.02 &  ---  &  0.21 & LINER     \\
               &OF-EB2$^*$& 0.09& ---   &  ---  & ---   & ---     \\

{\it Knot S4f} &          &     &       &       &       &       \\
               &MC-EMI$^*$& 0.19& -0.19 & -0.38 &  0.46 & LINER     \\
               &OF-EB2$^*$& --- & ---   &  ---  & ---   & ---     \\

               &          &     &       &       &       &         \\

\hline

\end{tabular}

\noindent
$^{a}$: [O {\sc iii}]$\lambda5007$; [O {\sc i}]$\lambda6300$; 
[N {\sc ii}]$\lambda6583$; [S {\sc ii}s]$\lambda\lambda$6716+6731;
[S {\sc ii}]/[S {\sc ii}] $\lambda6716$/$\lambda6731$.\\
Column 2: emission line components (Compon),
as in Table 3.\\
The values between parentheses are data with low S/N.

\end{table}


\clearpage

\begin{table}
\footnotesize \caption{Emission Lines of the Very Nucleus and South Nuclear Region
(at north-south direction, PA $=$ 180$^{\circ}$)}
\label{fluxs1}
\begin{tabular}{llcccccc}
\hline
\hline

Lines &Compon&        &Fluxes$^{a}$&      &        &        &         \\
      &     &Very Nuc.&0.2$''$S &0.4$''$S &0.6$''$S&0.8$''$S&1.0$''$S\\
      &     &N-B215+R238&N-B216+R237&N-B217+R236&N-B218+R235&N-B219+R234&N-B220+R233\\
      &     &         &         &         &        &        &      \\

\hline

  &         &         &         &         &        &        &      \\

H$\beta\lambda4861$
  & MC-EMI  & 0.19    & 0.33    & 0.17    &(0.28)  & 0.11   & 0.08 \\
  & OF-EB1  & 0.22    & 0.45    & 0.18    &(0.19)  & ---    & ---  \\
  &         &         &         &         &        &        &      \\

[O{\sc iii}]$\lambda5007$
  & MC-EMI  & 0.15    & 0.40    & 0.11    & 0.17   & 0.10   & 0.09 \\
  & OF-EB1  & 0.18    & 0.55    & 0.10    &(---)   & ---    & ---  \\
  &         &         &         &         &        &        &      \\

[O {\sc i}]$\lambda6300$
  & MC-EMI  & 0.80    & 0.95    & 0.68    &  0.20  & 0.27   & 0.29 \\
  & OF-EB1  & 0.75    & 0.80    & 0.85    &  0.40  & ---    & ---  \\
  &         &         &         &         &        &        &      \\

H$\alpha\lambda6563$
  & MC-EMI  & 1.20    & 0.85    & 0.40    & (---)  & 0.85   & 0.53  \\
  & OF-EB1  & 1.80    & 0.84    & 0.48    &  0.85  & ---    & ---   \\
  &         &         &         &         &        &        &       \\

[N {\sc ii}]$\lambda6583$
  & MC-EMI  &(0.80)   &(---)    &(---)    &  ---   & 0.36   & 0.33   \\
  & OF-EB1  & 1.47    &(---)    &(---)    &  0.37  & ---    & ---    \\
  &         &         &         &         &        &        &        \\

[S {\sc ii}]$\lambda6717$
  & MC-EMI  & 1.05    & 0.60    & 0.50    &  0.58  & 0.50   & 0.25   \\
  & OF-EB1  & 1.14    & 0.66    & 0.45    &  0.46  &(0.35)  & ---   \\
  &         &         &         &         &        &        &         \\

[S {\sc ii}]$\lambda6731$
  & MC-EMI  & 1.25    & 0.80    & 0.52    &  0.48  & 0.40   & 0.18   \\
  & OF-EB1  & 0.92    & 0.50    & 0.42    &  0.55  & 0.55   & ---    \\
  &         &         &         &         &        &        &         \\
  &         &         &         &         &        &        &         \\

H$\alpha$/H$\beta$
  & MC-EMI  & 6.3     & 3.0     & 3.1     &  ---   &  7.7   &  6.6    \\
  &         &         &         &         &        &        &         \\

FWHM-H$\alpha$ [km/s]
  & MC-EMI  &     85  &  80     &   87    &   95   &  110   &  155     \\

  &         &         &         &         &        &        &         \\

\hline

\end{tabular}

\noindent
$^{a}$: the fluxes are given in units of 10$^{-16}$ erg cm$^{-2}$ s$^{-1}$
(from GMOS/IFU-B600 spectroscopy).\\
Column 2: emission line components, as in Table 3.\\
The values between parentheses are data with low S/N.\\

\end{table}

\clearpage

\begin{table}
\footnotesize \caption{Emission Lines of the Very Nucleus and South Nuclear Region
(at PA $=$ 180$^{\circ}$). Cont.}
\label{fluxs2}
\begin{tabular}{llcccccc}
\hline
\hline

Lines &Compon&        &Fluxes$^{a}$&      &        &        &         \\
      &     &1.2$''$S &1.4$''$S &1.6$''$S &1.8$''$S&2.0$''$S&2.2$''$S\\
      &     &N-B221+R232&N-B222+R231&N-B223+R230&N-B224+R229&N-B225+R228&N-B226+R227\\
      &     &         &         &         &        &        &      \\

\hline

  &         &         &         &         &        &        &      \\

H$\beta\lambda4861$
  & MC-EMI  & 0.08    & 0.07    & 0.07    & 0.05   & 0.06   & 0.08 \\
  &         &         &         &         &        &        &      \\

[O{\sc iii}]$\lambda5007$
  & MC-EMI  & 0.09    & 0.11    & 0.07    & 0.08   & 0.08   & 0.10 \\
  &         &         &         &         &        &        &      \\

[O {\sc i}]$\lambda6300$
  & MC-EMI  & 0.25    & 0.15    & 0.16    & 0.12   & 0.13   & 0.12  \\
  &         &         &         &         &        &        &      \\

H$\alpha\lambda6563$
  & MC-EMI  & 0.44    & 0.38    & 0.33    & 0.27   & 0.18   & 0.25  \\
  &         &         &         &         &        &        &       \\

[N {\sc ii}]$\lambda6583$
  & MC-EMI  &  0.36   & 0.45    & 0.40    & 0.32   & 0.25   & 0.30   \\
  &         &         &         &         &        &        &        \\

[S {\sc ii}]$\lambda6717$
  & MC-EMI  & 0.30    & 0.20    & 0.22    & 0.19   & 0.13   & 0.15   \\
  &         &         &         &         &        &        &         \\

[S {\sc ii}]$\lambda6731$
  & MC-EMI  & 0.29    & 0.19    & 0.22    & 0.17   & 0.14   & 0.16   \\
  &         &         &         &         &        &        &         \\
  &         &         &         &         &        &        &         \\

H$\alpha$/H$\beta$
  & MC-EMI  & 5.5     & 5.4     & 4.7     &  5.4   &  3.0   &  3.1   \\
  &         &         &         &         &        &        &         \\

FWHM-H$\alpha$ [km/s]
  & MC-EMI  &  150    &  145    &  190    &  180   &  195   &  180    \\

  &         &         &         &         &        &        &         \\

\hline

\end{tabular}

\noindent
$^{a}$: the fluxes are given in units of 10$^{-16}$ erg cm$^{-2}$ s$^{-1}$
(from GMOS/IFU-B600 spectroscopy).\\
Column 2: emission line components, as in Table 3.\\
The values between parentheses are data with low S/N.\\

\end{table}

\clearpage

\begin{table}
\footnotesize \caption{Emission Lines of the North Nuclear Region
(at north-south direction, PA $=$ 00$^{\circ}$)}
\label{fluxn1}
\begin{tabular}{llccccc}
\hline
\hline

Lines &Compon&        &Fluxes$^{a}$&      &        &        \\
      &     &0.2$''$N &0.4$''$N &0.6$''$N &0.8$''$N&1.0$''$N\\
      &     &N-B214+R239&N-B213+R240&N-B212+R241&N-B211+R242&N-B210+R243 \\
      &     &         &         &         &        &        \\

\hline

  &         &         &         &         &        &         \\

H$\beta\lambda4861$
  & MC-EMI  & 0.15    & 0.12    & 0.13    & 0.08   & 0.13     \\
  &         &         &         &         &        &          \\

[O{\sc iii}]$\lambda5007$
  & MC-EMI  & 0.13    & 0.11    & 0.08    & 0.05   & 0.11     \\
  &         &         &         &         &        &            \\

[O {\sc i}]$\lambda6300$
  & MC-EMI  & 0.78    & 0.66    & 0.48    & 0.28   & 0.23      \\
  &         &         &         &         &        &            \\

H$\alpha\lambda6563$
  & MC-EMI  & 1.70    & 0.95    & 1.15    & 0.55   & 0.80       \\
  & OF-EB1  & ---     & ---     & ---     & 0.40   & 0.50       \\
  &         &         &         &         &        &             \\

[N {\sc ii}]$\lambda6583$
  & MC-EMI  & (---)   &(---)    & 0.40    & 0.35   & 0.61        \\
  & OF-EB1  & ---     & ---     & ---     &(---)   &(---)        \\
  &         &         &         &         &        &             \\

[S {\sc ii}]$\lambda6717$
  & MC-EMI  & 1.20    & 0.78    & 0.35    & 0.55   & 0.30        \\
  & OF-EB1  & 0.98    & 0.93    & 0.34    & 0.11   & 0.20        \\
  &         &         &         &         &        &             \\

[S {\sc ii}]$\lambda6731$
  & MC-EMI  & 1.45    & 0.85    & 0.48    & 0.50   & 0.40        \\
  & OF-EB1  & 0.70    & 0.61    & 0.25    & 0.10   & 0.10        \\
  &         &         &         &         &        &              \\
  &         &         &         &         &        &              \\

H$\alpha$/H$\beta$
  & MC-EMI  & 11.3    & 8.0     & 8.8     & 7.0    & 6.2         \\
  &         &         &         &         &        &              \\

FWHM-H$\alpha$  [km/s]
  & MC-EMI  &  85     &  95     & 120     & 115    &  125        \\

  &         &         &         &         &        &              \\

\hline

\end{tabular}

\noindent
$^{a}$: the fluxes are given in units of 10$^{-16}$ erg cm$^{-2}$ s$^{-1}$
(from GMOS/IFU-B600 spectroscopy).\\
Column 2: emission line components, as in Table 3.\\
The values between parentheses are data with low S/N.\\

\end{table}

\clearpage

\begin{table}
\footnotesize \caption{Emission Lines of the  North Nuclear Region
(at PA $=$ 00$^{\circ}$). Contin.}
\label{fluxn2}
\begin{tabular}{llcccccc}
\hline
\hline

Lines &Compon&        &Fluxes$^{a}$&      &        &        &         \\
      &     &1.2$''$N &1.4$''$N &1.6$''$N &1.8$''$N&2.0$''$N&2.2$''$N\\
      &     &N-B209+R244&N-B208+R245&N-B207+R246&N-B206+R247&N-B205+R248&N-B204+R249\\
      &     &         &         &         &        &        &      \\

\hline

  &         &         &         &         &        &        &      \\

H$\beta\lambda4861$
  & MC-EMI  &(0.15)   & 0.20    & 0.21    & 0.15   &(---)   &(---) \\
  &         &         &         &         &        &        &      \\

[O{\sc iii}]$\lambda5007$
  & MC-EMI  &(0.10)   & 0.22    & 0.16    & 0.12   &(---)   &(---) \\
  &         &         &         &         &        &        &      \\

[O {\sc i}]$\lambda6300$
  & MC-EMI  & 0.16    & 0.13    & 0.15    & 0.10   & 0.15   & 0.17  \\
  & OF-EB1  & 0.14    & 0.10    & ---     & 0.07   & ---    & ---  \\
  &         &         &         &         &        &        &      \\

H$\alpha\lambda6563$
  & MC-EMI  & 1.68    & 1.64    & 1.18    & 0.70   & 0.44   & 0.31  \\
  & OF-EB1  & ---     & ---     & 0.18    & ---    & ---    & ---   \\
  &         &         &         &         &        &        &       \\

[N {\sc ii}]$\lambda6583$
  & MC-EMI  &  0.95   & 1.23    & 1.19    & 0.61   & 0.45   & 0.40   \\
  &         &         &         &         &        &        &        \\

[S {\sc ii}]$\lambda6717$
  & MC-EMI  & 0.26    & 0.33    & 0.44    & 0.33   & 0.18   & 0.20   \\
  & OF-EB1  & 0.15    & 0.22    & 0.10    & 0.08   & ---    & ---   \\
  &         &         &         &         &        &        &         \\

[S {\sc ii}]$\lambda6731$
  & MC-EMI  & 0.35    & 0.35    & 0.30    & 0.24   & 0.16   & 0.25   \\
  & OF-EB1  & 0.10    & 0.10    & 0.07    & 0.10   & ---    & ---    \\
  &         &         &         &         &        &        &         \\
  &         &         &         &         &        &        &         \\

H$\alpha$/H$\beta$
  & MC-EMI  & (11.2)  & 8.2     & 5.6     & 4.7    & ---    & ---     \\
  &         &         &         &         &        &        &         \\

FWHM-H$\alpha$  [km/s]
  & MC-EMI  &         &         &         &        &        &         \\

  &         &         &         &         &        &        &         \\

\hline

\end{tabular}

\noindent
$^{a}$: the fluxes are given in units of 10$^{-16}$ erg cm$^{-2}$ s$^{-1}$
(from GMOS/IFU-B600 spectroscopy).\\
Column 2: emission line components, as in Table 3.\\
The values between parentheses are data with low S/N.\\

\end{table}

\clearpage

\begin{table}
\footnotesize \caption{Emission Line Ratios of the Very Nucleus and
South Nuclear Region (at north-south direction, PA $=$ 180$^{\circ}$)}
\label{elrs4}
\begin{tabular}{lcccccc}
\hline \hline

               &       &       &       &       &       &      \\
Regions        &Compon& lg[O{\sc iii}]/H$\beta$$^{a}$&lg[O{\sc i}]/H$\alpha^{a}$ 
& lg[N{\sc ii}]/H$\alpha$$^{a}$  & lg[S{\sc ii}s]/H$\alpha$$^{a}$ & Spectral Type \\
\hline
               &       &       &       &       &       &       \\

{\it Very Nucleus}&       &     &       &       &       &       \\
               &MC-EMI    &-0.09& -0.18 &(-0.18)&  0.05 & LINER     \\

{\it 0.2$''$ S}&          &     &       &       &       &       \\
               &MC-EMI    & 0.07&  0.01 &  ---  &  0.25 & LINER     \\

{\it 0.4$''$ S}&          &     &       &       &       &       \\
               &MC-EMI    &-0.18&  0.25 &  ---  &  0.42 & LINER     \\

{\it 0.6$''$ S}&          &     &       &       &       &       \\
               &MC-EMI    &-0.18& -0.33 &  ---  &  0.05 & LINER     \\

{\it 0.8$''$ S}&          &     &       &       &       &       \\
               &MC-EMI    &-0.04& -0.43 & -0.37 &  0.03 & LINER     \\

{\it 1.0$''$ S}&          &     &       &       &       &       \\
               &MC-EMI    & 0.55& -0.26 & -0.20 & -0.05 & LINER     \\

{\it 1.2$''$ S}&          &     &       &       &       &       \\
               &MC-EMI    & 0.05& -0.25 & -0.09 &  0.14 & LINER     \\

{\it 1.4$''$ S}&          &     &       &       &       &       \\
               &MC-EMI    & 0.19& -0.35 &  0.07 &  0.02 & LINER     \\

{\it 1.6$''$ S}&          &     &       &       &       &       \\
               &MC-EMI    & 0.00& -0.31 &  0.31 &  0.13 & LINER     \\

{\it 1.8$''$ S}&          &     &       &       &       &       \\
               &MC-EMI    & 0.20& -0.35 &  0.07 &  0.13 & LINER     \\

{\it 2.0$''$ S}&          &     &       &       &       &       \\
               &MC-EMI    & 0.06& -0.14 &  0.14 &  0.18 & LINER     \\

{\it 2.2$''$ S}&          &     &       &       &       &       \\
               &MC-EMI    & 0.05& -0.28 &  0.11 &  0.12 & LINER     \\

               &          &     &       &       &       &         \\

\hline

\end{tabular}

\noindent
$^{a}$: [O {\sc iii}]$\lambda5007$; [O {\sc i}]$\lambda6300$; 
[N {\sc ii}]$\lambda6583$; [S {\sc ii}s]$\lambda\lambda$6716+6731.\\
Column 2: emission line components (Compon),
as in Table 3.\\
The values between parentheses are data with low S/N.

\end{table}

\clearpage

\begin{table}
\footnotesize \caption{Emission Line Ratios of the North Nuclear
Region (at north-south direction, PA $=$ 00$^{\circ}$)}
\label{elrn}
\begin{tabular}{lcccccc}
\hline \hline

               &       &       &       &       &       &      \\
Regions        &Compon& lg[O{\sc iii}]/H$\beta$$^{a}$&lg[O{\sc i}]/H$\alpha^{a}$ 
& lg[N{\sc ii}]/H$\alpha$$^{a}$  & lg[S{\sc ii}s]/H$\alpha$$^{a}$ & Spectral Type \\
\hline
               &       &       &       &       &       &       \\

{\it 0.2$''$ N}&          &     &       &       &       &       \\
               &MC-EMI    &-0.07& -0.34 &  ---  &  0.19 & LINER     \\

{\it 0.4$''$ N}&          &     &       &       &       &       \\
               &MC-EMI    &-0.13& -0.12 &  ---  &  0.26 & LINER     \\

{\it 0.6$''$ N}&          &     &       &       &       &       \\
               &MC-EMI    &-0.21& -0.38 &  ---  & -0.10 & LINER     \\

{\it 0.8$''$ N}&          &     &       &       &       &       \\               &MC-EMI    &-0.25& -0.44 & -0.56 & -0.13 & LINER     \\
               &MC-EMI    &-0.20& -0.29 &  0.21 &  0.28 & LINER     \\

{\it 1.0$''$ N}&          &     &       &       &       &       \\
               &MC-EMI    &-0.07& -0.25 & -0.12 & -0.05 & LINER     \\

{\it 1.2$''$ N}&          &     &       &       &       &       \\
               &MC-EMI    &-0.12& -1.00 & -0.25 & -0.44 & LINER     \\

{\it 1.4$''$ N}&          &     &       &       &       &       \\
               &MC-EMI    & 0.04& -1.10 & -0.12 & -0.38 & LINER     \\

{\it 1.6$''$ N}&          &     &       &       &       &       \\
               &MC-EMI    &-0.12& -0.90 &  0.00 & -0.20 & LINER     \\

{\it 1.8$''$ N}&          &     &       &       &       &       \\
               &MC-EMI    &-0.09& -0.85 & -0.06 & -0.12 & LINER     \\

{\it 2.0$''$ N}&          &     &       &       &       &       \\
               &MC-EMI    &-0.10& -0.47 &  0.00 & -0.11 & LINER     \\

{\it 2.2$''$ N}&          &     &       &       &       &       \\
               &MC-EMI    & --- & -0.24 &  0.11 &  0.13 & LINER     \\

               &          &     &       &       &       &         \\

\hline

\end{tabular}

\noindent
$^{a}$: [O {\sc iii}]$\lambda5007$; [O {\sc i}]$\lambda6300$; 
[N {\sc ii}]$\lambda6583$; [S {\sc ii}s]$\lambda\lambda$6716+6731.\\
Column 2: emission line components (Compon),
as in Table 3.\\
The values between parentheses are data with low S/N.

\end{table}


\clearpage

\begin{figure*}
\vspace{12.0 cm}
\begin{tabular}{c}
\includegraphics{fig1z.ps}\cr
\end{tabular}
\vspace{8.0 cm}
\caption {V broad band image of the main body and the faint tails of Mrk 231
(obtained at the 2.5 NOT telescope, La Palma Spain) showing the mosaic field
observed with Gemini+GMOS.
}
\label{f1not}
\end{figure*}

\clearpage

\begin{figure*}
\vspace{12.0 cm}
\begin{tabular}{c}
\includegraphics{fig2az.ps} \cr
\includegraphics{fig2bz.ps} \cr
\end{tabular}
\vspace{6.0 cm}
\caption {
Example of GMOS spectra with multi or different emission and
absorption line components/systems.
The scales of flux are given in units of [erg $\times$ cm$^{-2}$ $\times$
s$^{-1}$ $\times$ \AA$^{-1}$ $\times$ 10$^{-16}$].
}
\label{fig2}
\end{figure*}

\clearpage

\begin{figure*}
\vspace{12.0 cm}
\begin{tabular}{c}
\includegraphics{fig2cz.ps} \cr
\end{tabular}
\vspace{9.0 cm}
\addtocounter{figure}{-1}
\caption {
Continued.
}
\label{fig2c}
\end{figure*}

\clearpage

\begin{figure*}
\vspace{12.0 cm}
\begin{tabular}{c}
\includegraphics{fig3z.ps}\cr
\end{tabular}
\vspace{8.0 cm}
\caption {Combined 1D KPNO and HST/FOS  spectra of Mrk 231.
 }
\label{fig3}
\end{figure*}

\clearpage

\begin{figure*}
\vspace{12.0 cm}
\begin{tabular}{c}
\includegraphics{fig4az.ps} \cr
\includegraphics{fig4bz.ps} \cr
\end{tabular}
\vspace{6.0 cm}
\caption {Decoupled QSO and host galaxy spectra of Mrk 231.
These spectra were obtained from the decoupling of the Gemini GMOS IFU
B600 spectra, using the new technique described by Sanchez et al.
(2006a,b, 2004) [see the text].
 }
\label{fig4}
\end{figure*}

\clearpage

\begin{figure*}
\vspace{12.0 cm}
\begin{tabular}{c}
\includegraphics{fig5az.ps} \cr
\includegraphics{fig5bz.ps} \cr
\end{tabular}
\vspace{6.0 cm}
\caption {
Detailed QSO and host galaxy nuclear spectra, of Mrk 231, for
selected wavelength regions (using the B600 grating).
In the panel 5(b) and for the host galaxy the y-axis was
expanded (in order to show the weak narrow Na ID absorption
lines)
}
\label{fig5}
\end{figure*}

\clearpage

\begin{figure*}
\vspace{12.0 cm}
\begin{tabular}{c}
\includegraphics{fig5cz.ps} \cr
\end{tabular}
\vspace{9.0 cm}
\addtocounter{figure}{-1}
\caption {
Continued.
}
\label{fig5c}
\end{figure*}

\clearpage

\begin{figure*}
\vspace{12.0 cm}
\begin{tabular}{cc}
\includegraphics{fig6az.ps}&
\includegraphics{fig6bz.ps} \cr
\includegraphics{fig6cz.ps}&
\includegraphics{fig6dz.ps} \cr
\end{tabular}
\vspace{6.0 cm}
\caption {
Sequence of individual GMOS spectra --at the blue wavelength region-- of the
very nucleus and the nuclear region of Mrk 231 -at PA = 00$^{\circ}$-  showing the presence of
the strong blue (and red) continuum.
The blue component was found in all the south region.
The scales of flux are given in units of [erg $\times$ cm$^{-2}$ $\times$
s$^{-1}$ $\times$ \AA$^{-1}$ $\times$ 10$^{-16}$].
 }
\label{fig6}
\end{figure*}

\clearpage

\begin{figure*}
\vspace{12.0 cm}
\begin{tabular}{cc}
\includegraphics{fig6ez.ps}&
\includegraphics{fig6fz.ps} \cr
\includegraphics{fig6gz.ps}&
\includegraphics{fig6hz.ps} \cr
\end{tabular}
\vspace{6.0 cm}
\addtocounter{figure}{-1}
\caption {Continued
}
\label{fig6c}
\end{figure*}

\clearpage

\begin{figure*}
\vspace{12.0 cm}
\begin{tabular}{cc}
\includegraphics{fig7az.ps}&
\includegraphics{fig7bz.ps} \cr
\includegraphics{fig7cz.ps}&
\includegraphics{fig7dz.ps} \cr
\end{tabular}
\vspace{6.0 cm}
\caption {
Sequence of individual GMOS spectra --at the H$\beta$ wavelength region-- of the
very nucleus and the nuclear region of Mrk 231 -at PA = 00$^{\circ}$-  showing the presence of
the strong blue (and red) continuum.
The blue component was found in all the south region.
The scales of flux are given in units of [erg $\times$ cm$^{-2}$ $\times$
s$^{-1}$ $\times$ \AA$^{-1}$ $\times$ 10$^{-16}$].
 }
\label{fig7}
\end{figure*}

\clearpage

\begin{figure*}
\vspace{12.0 cm}
\begin{tabular}{cc}
\includegraphics{fig7ez.ps}&
\includegraphics{fig7fz.ps} \cr
\includegraphics{fig7gz.ps}&
\includegraphics{fig7hz.ps} \cr
\end{tabular}
\vspace{6.0 cm}
\addtocounter{figure}{-1}
\caption {Continued
}
\label{fig7c}
\end{figure*}

\clearpage

\begin{figure*}
\vspace{12.0 cm}
\begin{tabular}{cc}
\includegraphics{fig8az.ps}&
\includegraphics{fig8bz.ps} \cr
\end{tabular}
\vspace{8.0 cm}
\caption {GMOS map and contour of the strong blue (and red) continuum
components.
 }
\label{fig8}
\end{figure*}

\clearpage

\begin{figure*}
\vspace{12.0 cm}
\begin{tabular}{c}
\includegraphics{fig9z.ps} \cr
\end{tabular}
\vspace{6.0 cm}
\caption {
Selected GMOS spectra of the very nucleus  and nuclear region
of Mrk 231 --for the red wavelength--
showing the IR Ca {\sc ii} triplet. These spectra were obtained with
the best GMOS-IFU spectral resolution (and in our Gemini observing run
with the best seeing/spatial resolution).
 }
\label{fig9}
\end{figure*}

\clearpage

\begin{figure*}
\vspace{12.0 cm}
\begin{tabular}{c}
\includegraphics{fig10z.ps} \cr
\end{tabular}
\vspace{6.0 cm}
\caption {
Sequence of 3D Gemini GMOS IFU spectra, along the north south 
direction (at PA $=$ 00$^{\circ}$) showing  the extended nature
of the Na ID$\lambda$5889-95 BAL system I.
The scales of flux are given in units of [erg $\times$ cm$^{-2}$ $\times$
s$^{-1}$ $\times$ \AA$^{-1}$].
}
\label{fig10}
\end{figure*}

\clearpage

\begin{figure*}
\vspace{12.0 cm}
\begin{tabular}{cc}
\includegraphics{fig11az.ps}& 
\includegraphics{fig11bz.ps} \cr
\includegraphics{fig11cz.ps}&
\includegraphics{fig11dz.ps} \cr
\includegraphics{fig11ez.ps}& 
\includegraphics{fig11fz.ps} \cr
\end{tabular}
\vspace{7.0 cm}
\caption {
GMOS maps of BALs.
 }
\label{fig11}
\end{figure*}

\clearpage

\begin{figure*}
\vspace{12.0 cm}
\begin{tabular}{c}
\includegraphics{fig12az.ps}\cr
\includegraphics{fig12bz.ps}\cr
\includegraphics{fig12cz.ps}\cr
\end{tabular}
\vspace{8.0 cm}
\caption {
Na ID BAL I, II and III systems:
(a) High resolution (R831) spectrum of the BAL I and II systems;
(b) Medium resolution (B600) spectrum of the weak BAL III system;
(c) Light curve variability of the Na ID BAL III system
(for the fall, between 1988 and 2005) .
}
\label{fig12}
\end{figure*}

\clearpage

\begin{figure*}
\vspace{12.0 cm}
\begin{tabular}{c}
\includegraphics{fig13az.ps} \cr
\includegraphics{fig13bz.ps} \cr
\end{tabular}
\vspace{6.0 cm}
\caption {
HST-WFPC2 F439W  B image  (a) and  residuals--image (b) of Mrk 231
are depicted. Which shows the main shells and their knots (see the text).
These images were adapted from Lipari et al. (2005a).
 }
\label{fig13}
\end{figure*}

\clearpage

\begin{figure*}
\vspace{12.0 cm}
\begin{tabular}{cc}
\includegraphics{fig14az.ps}&
\includegraphics{fig14bz.ps} \cr
\includegraphics{fig14cz.ps}&
\includegraphics{fig14dz.ps} \cr
\end{tabular}
\vspace{6.0 cm}
\caption {
Gemini+GMOS mosaic maps (3.5$'' \times$ 9$''$) of the continuum (a, b)
and of the narrow band of the redshifted H$\alpha$, H$\beta$,
[O {\sc iii}]$\lambda$5007 and [S {\sc ii}]$\lambda$6716+31 (c, d, e, f)
showing the main knots in the shell S1 (see the text).
The nucleus (in each GMOS mosaic maps) is positioned at
$\sim\Delta\alpha =$ 0$''$, and $\sim\Delta\delta =$ +2$''$
}
\label{fig14}
\end{figure*}

\clearpage

\begin{figure*}
\vspace{12.0 cm}
\begin{tabular}{cc}
\includegraphics{fig14ez.ps}&
\includegraphics{fig14fz.ps} \cr
\end{tabular}
\vspace{6.0 cm}
\addtocounter{figure}{-1}
\caption {Continued
}
\label{fig14c}
\end{figure*}

\clearpage

\begin{figure*}
\vspace{12.0 cm}
\begin{tabular}{c}
\includegraphics{fig15az.ps}\cr
\includegraphics{fig15bz.ps}\cr
\includegraphics{fig15cz.ps}\cr
\end{tabular}
\vspace{6.0 cm}
\caption {
Examples of Gemini + GMOS 3D spectra  in the main knots
of the 4 external supergiant shells of Mrk 231
(showing several OF components).
The scales of flux are given in units of [erg $\times$ cm$^{-2}$ $\times$
s$^{-1}$ $\times$ \AA$^{-1}$ $\times$ 10$^{-16}$].
}
\label{fig15}
\end{figure*}

\clearpage

\begin{figure*}
\vspace{12.0 cm}
\begin{tabular}{c}
\includegraphics{fig15dz.ps}\cr
\end{tabular}
\vspace{6.0 cm}
\addtocounter{figure}{-1}
\caption {Continued
}
\label{fig15c}
\end{figure*}

\clearpage

\begin{figure*}
\vspace{12.0 cm}
\begin{tabular}{c}
\includegraphics{fig16z.ps}\cr
\end{tabular}
\vspace{6.0 cm}
\caption {GMOS spectra showing Wolf Rayet feature,
in the knot K14-East, at the more external shell S1 (see the text).
}
\label{fig16}
\end{figure*}

\clearpage

\begin{figure*}
\vspace{12.0 cm}
\begin{tabular}{c}
\includegraphics{fig17z.ps}\cr
\end{tabular}
\vspace{6.0 cm}
\caption {Superposition of the WHT+INTEGRAL H$\alpha$ velocity field map
for the ionized gas (of the central region of Mrk 231) and the
HST WFPC2-I contour image.
}
\label{fig17-dkinha}
\end{figure*}

\clearpage

\begin{figure*}
\vspace{12.0 cm}
\begin{tabular}{cc}
\includegraphics{fig18az.ps} &
\includegraphics{fig18bz.ps} \cr
\end{tabular}
\vspace{6.0 cm}
\caption {GMOS maps of the emission line ratios: [N {\sc ii}]/H$\alpha$
and [S {\sc ii}]/H$\alpha$.
The nucleus (in each GMOS mosaic maps) is positioned at
$\sim\Delta\alpha =$ 0$''$, and $\sim\Delta\delta =$ +2$''$.
}
\label{fig18}
\end{figure*}

\clearpage

\begin{figure*}
\vspace{12.0 cm}
\begin{tabular}{c}
\includegraphics{fig19az.ps} \cr
\includegraphics{fig19bz.ps} \cr
\end{tabular}
\vspace{6.0 cm}
\caption {
Emission line ratio diagnostic diagram of:
(a) the main knots of the 4 external supergiant bubbles
(grey symbols are knots located in the south-west area).
(b) for a sequence of the  GMOS spectra, for the very nucleus
and the nuclear region (with step of 0.2$''$ and at the PA  $=$ 00).
The fill symbols show the values for r $<$ 0.6$''$.
The main areas of these figures were adapted from Heckman et al.
(1990; their Fig. 14).
}
\label{fig19}
\end{figure*}

\clearpage

\begin{figure*}
\vspace{12.0 cm}
\begin{tabular}{cc}
\includegraphics{fig20az.ps}&
\includegraphics{fig20bz.ps} \cr
\includegraphics{fig20cz.ps}&
\includegraphics{fig20dz.ps} \cr
\end{tabular}
\vspace{6.0 cm}
\caption {GMOS velocity field maps (a,b,c,d) for emission and
stellar, ISM absorption lines. Plus a rotation curve, derived
from the stellar kinematics.
In the GMOS mosaic maps (a and b) the nucleus is positioned at
$\sim\Delta\alpha =$ 0$''$, and $\sim\Delta\delta =$ +2$''$.
For the GMOS one frame maps (c and d) the nucleus is positioned at
$\sim\Delta\alpha =$ 0$''$, and $\sim\Delta\delta =$ 0$''$.
The white circles show the nuclear region where
the narrow emission lines were not detected.
}
\label{fig20}
\end{figure*}

\clearpage

\begin{figure*}
\vspace{12.0 cm}
\begin{tabular}{c}
\includegraphics{fig20ez.ps} \cr
\end{tabular}
\vspace{9.0 cm}
\addtocounter{figure}{-1}
\caption {
Continued.
}
\label{fig20c}
\end{figure*}

\end{document}